\documentclass[11pt,onecolumn,draftcls]{IEEEtran}

\def\comment#1{}
\usepackage{graphicx,amsfonts,amsmath,amssymb,xcolor,amsthm}
\usepackage{algorithm,algorithmic,bm,color}
\usepackage{multirow}
\usepackage{hyperref}
\usepackage[noadjust]{cite}
\usepackage{adjustbox}
\usepackage{makecell}
\def\comment#1{}
\newcommand{\argmin}{\arg\!\min}

\newcommand{\Dmat}{{\bf D}}
\newcommand{\Emat}[0]{{{\bf E}}}
\newcommand{\Fmat}[0]{{{\bf F}}}

\newcommand{\Imat}{{\bf I}}

\newcommand{\Mmat}[0]{{{\bf M}}}

\newcommand{\Qmat}[0]{{{\bf Q}}}

\newcommand{\Xmat}{{\bf X}}
\newcommand{\Ymat}[0]{{{\bf Y}}}

\newcommand{\av}{\boldsymbol{a}}

\newcommand{\cv}{{\boldsymbol{c}}}

\newcommand{\ev}[0]{{\boldsymbol{e}}}

\newcommand{\uv}[0]{{\boldsymbol{u}}}
\newcommand{\vv}{\boldsymbol{v}}

\newcommand{\xv}{\boldsymbol{x}}
\newcommand{\yv}{\boldsymbol{y}}

\newcommand{\Sigmamat}{\boldsymbol{\Sigma}}

\newcommand{\Phimat}{\boldsymbol{\Phi}}
\newcommand{\Psimat}{\boldsymbol{\Psi}}

\newcommand{\muv}{\boldsymbol{\mu}}

\newcommand{\ts}{^{\top}}
\newcommand{\inv}{^{-1}}
\newcommand{\ie}{{\em i.e.}}
\newcommand{\eg}{{\em e.g.}}
\newcommand{\etc}{{\em etc}}

\newtheorem{theorem}{Theorem}

\newtheorem{corollary}{Corollary}

\newcommand{\gos}{\rightarrow}

\begin{document}
\setlength{\parskip}{.02in}

\title{Snapshot Compressive Imaging: \\Principle, Implementation, Theory, \\ Algorithms and Applications}	
	
%
\author{
	\IEEEauthorblockN{Xin Yuan, {\em Senior Member, IEEE,}	 David J. Brady, {\em Fellow, IEEE} \\ and Aggelos K. Katsaggelos, {\em Fellow, IEEE}} 
\thanks{Xin Yuan is with Bell Labs, 600 Mountain Avenue, Murray Hill, NJ 07974, USA, Email: xyuan@bell-labs.com.   \newline David J. Brady is with College of Optical Sciences, University of Arizona, Tucson, Arizona 85719, USA, Email: djbrady@arizona.edu. \newline Aggelos K. Katsaggelos is with Department of Electrical and Computer Engineering, Northwestern University, Evanston, IL 60208, USA, Email: aggk@eecs.northwestern.edu}
	}	
%

\maketitle
\begin{abstract}
Capturing high-dimensional (HD) data is a long-term challenge in signal processing and related fields. Snapshot compressive imaging (SCI) uses a two-dimensional (2D) detector to capture HD ($\ge3$D) data in a {\em snapshot} measurement.  Via novel optical designs, the 2D detector samples the HD data in a {\em compressive} manner; following this, algorithms are employed to reconstruct the desired HD data-cube. SCI has been used in hyperspectral imaging, video, holography, tomography, focal depth imaging, polarization imaging, microscopy, \etc.~Though the hardware has been investigated for more than a decade, the theoretical guarantees have only recently been derived. Inspired by deep learning, various deep neural networks have also been developed to reconstruct the HD data-cube in spectral SCI and video SCI. This article reviews recent advances in SCI hardware, theory and algorithms, including both optimization-based and deep-learning-based algorithms. Diverse applications and the outlook of SCI are also discussed.
\end{abstract}
	

\section{Introduction \label{Sec:intro}}
We live in a high-dimensional (HD) world. Sensing and capturing the HD data (signal) around us is the first step to perception and understanding. Recent advances in artificial intelligence and robotics have produced an unprecedented demand for HD data capture and processing. Unfortunately, most physical optical, x-ray, radar and acoustic sensors consist of 2D arrays. Sensors that directly perform 3D (and larger) data acquisition generally do not exist. 


To address this challenge, snapshot compressive imaging (SCI) utilizes a 2D detector to capture HD ($\ge3$D) data. Equipped with advanced compressive sensing (CS) algorithms~\cite{Candes06_Robust,Donoho06_CS}, SCI has demonstrated promising results on hyperspectral~\cite{Gehm07_DDCASSI,Wagadarikar08_CASSI,Wagadarikar09_vCASSI,Lin14_ACMTG,Yuan15_JSTSP_CASSI_SI}, temporal~\cite{Hitomi11_ICCV_videoCS,Reddy11_CVPR_P2C2,Llull13_OE_CACTI,Gao14_Nature_fast,Yuan16BOE}, volumetric~\cite{brady2015compressive,Qiao19_OCT}, holographic~\cite{Brady09_ComHolo,Brady18_CPRL_omHolo,Wang17_OE_HoloVideo}, light field~\cite{He20_Lightfield_OE}, polarization~\cite{Tsai15OE}, high dynamic range~\cite{Nayar_2000_CVPR_HDR,Nayar_03_ICCV_HDRSLM,Yuan_16_OE}, focal~\cite{Llull15Optica}, spectral-temporal~\cite{Tsai15OL}, spectral-polarization~\cite{Tsai13_AO_SpecPol}, spatial-temporal~\cite{Qiao2020_CACTI}, range-temporal~\cite{Sun16OE,Sun17OE}, holographic-temporal~\cite{Wang17_OE_HoloVideo}  and focal-temporal~\cite{Yuan14CVPR} data. These systems utilize novel optical designs to capture the compressed 2D measurement and then employ CS algorithms~\cite{Bioucas-Dias2007TwIST,Yuan16ICIP_GAP,Yang14GMM,Yang14GMMonline,Rajwade13,Yuan16AO,Yuan2020_CVPR_PnP,Qiao2020_APLP} to reconstruct the HD data. This leads to a {\em hardware encoder plus a software decoder} system, which enjoys the advantages of low bandwidth/memory requirements, fast acquisition speed and potentially low cost and low power. Here, the ``hardware encoder" denotes the optical imaging system of the SCI while the ``software decoder" denotes the reconstruction algorithm. In a nutshell, SCI systems modulate the HD data with a higher speed than the capture rate of the camera. With knowledge of the modulation, HD data can be reconstructed from each compressed measurement by using advanced CS algorithms.
This article will give a conceptual tutorial on SCI and then provide a review of recent advances, including the hardware design, theoretical analysis, algorithm developments and applications.

\subsection{Motivation and Challenges \label{Sec:Motivation}}
As mentioned above, a number of systems summarized in Table~\ref{Tab:SCI_system} demonstrate the feasibility of the use of SCI to capture high-speed HD data. In general, SCI belongs to the scientific area computational imaging~\cite{Mait18_AOP_CI,Altmann18Science}, a multidisciplinary topic, in which optics, computation and signal processing come together.

Though related to CS, the forward model of SCI is significantly different from the global random transformation-based CS, \eg, the single-pixel camera regime~\cite{Duarte08SPM} in Fig.~\ref{Fig:forward} (top). It is more closely related to sparse MRI~\cite{lustig2007sparse},
which is also a tomographic approach but deals with a different sampling structure. In contrast with global transformations, both SCI and sparse MRI incorporate compact forward model support to improve conditioning and a physics-based sampling structure. 

The theoretical basis of SCI has been recently addressed in~\cite{Jalali19TIT_SCI}. 
In parallel to this theoretical advance, deep learning based algorithms~\cite{Iliadis18DSPvideoCS,Lucas18SPM,Ma19ICCV,Miao19ICCV,Yuan2020_CVPR_PnP,Qiao2020_APLP,Iliadis_DSP2020,Meng2020_OL_SHEM,Meng20ECCV_TSAnet,Cheng20ECCV_Birnat} have recently shown promise in end-to-end SCI systems.  



From the signal processing perspective, deep learning has brought new challenges for SCI reconstruction, such as: $i$) interpretability of convolutional neural networks (CNNs) for SCI reconstruction and other inverse problems, $ii$) efficient algorithm design using (probably pre-trained or training-free) deep learning networks, and $iii$) convergence analysis of algorithms using deep learning networks. In this article, we aim to address the first two challenges by detailing the deep-learning based reconstruction frameworks in Sec.~\ref{Sec:bridge}. Convergence will also be briefly discussed, and this is an important direction for future research. 
With the advances of deep learning, real-time reconstruction is anticipated. Based on this, we expect to see more and more SCI systems being used in our daily life. In addition, machine-learning based task driven SCI systems will be developed not only for reconstruction but also for anomaly detection, pattern recognition and other tasks.


\subsection{Organization of This Article}
In Sec. \ref{Sec:Sys} we provide the underlying principle of SCI and briefly describe some exemplar SCI systems.
The mathematical model of SCI is derived in Sec.~\ref{Sec:Model} and in Sec.~\ref{Sec:Theory} we present the theoretical guarantees. Regularization based conventional optimization algorithms, shallow-learning algorithms and deep learning algorithms are introduced in Sec. \ref{Eq:regu_opt}, \ref{Sec:Shallow_Learn}, and \ref{Sec:Deep}, respectively. In Sec. \ref{Sec:bridge} we provide recent research results for bridging the gap between deep learning and traditional (iterative) optimization algorithms.
Diverse applications and the outlook of SCI systems are discussed in Sec.~\ref{Sec:Appli} and Sec.~\ref{Sec:Outlook}, respectively.

\section{Imaging Systems of SCI \label{Sec:Sys}}
In this article, we will use SCI for 3D signals to demonstrate the ideas, theory, and algorithms, with video SCI and spectral SCI as two representative applications. The described principles can be readily extended to other HD SCI systems.

\begin{figure}[h!]
	\centering
	{\includegraphics[width=1\columnwidth]{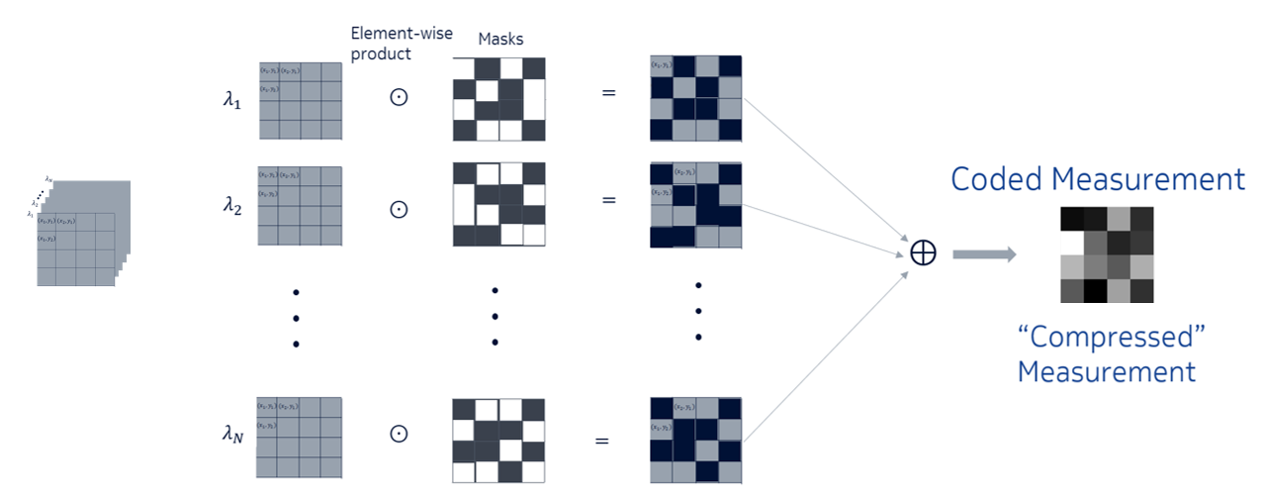}}
	\caption{Encoding process of SCI. Left: the desired 3D data-cube, shown here as $(x,y,\lambda)$; right: each frame (a 2D image corresponding to a specific wavelength) is modulated by a different mask and then integrated into the 2D ``compressed" measurement.}
	\label{Fig:sci_encoder}
\end{figure}

\begin{figure}[!]
	\centering
	{\includegraphics[width=1\columnwidth]{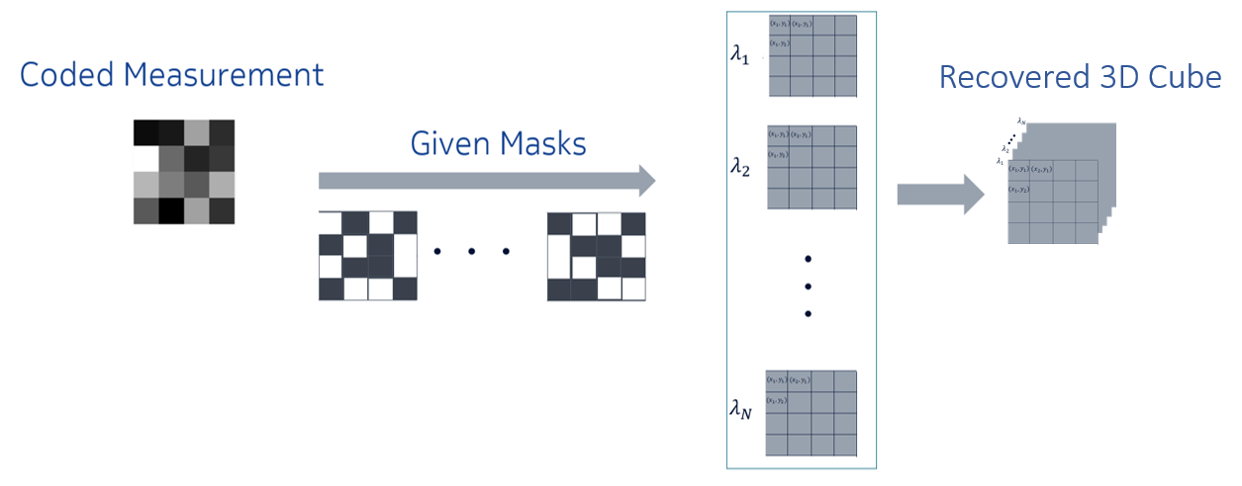}}
	\caption{Decoding process of SCI. The ``coded measurement" captured by the SCI system is sent to the algorithm (either optimization based, or deep learning based) along with the masks to recover the data-cube.}
	\label{Fig:sci_decoder}
\end{figure}

\begin{table}[!htbp]
	\caption{Different SCI systems to captured diverse HD data-cube. LCoS: Liquid Crystal on Silicon; DMD: Digital Micromirror Device. $(x,y)$: 2D spatial coordinates; $\lambda$: wavelength (spectrum); $t$: time; $z$: depth; $p$: polarization; $d$: dynamic range; $(x,y,z)$: 3D tomography or holography; $(x,y)+z$: 2D image plus depth map.}
	\centering
	{
		\begin{tabular}{|l|l|l|}
			\hline
			Data-cube& Modulation method & Reference \\
			\hline
			$(x,y,\lambda)$ & \makecell[l]{Fixed mask + disperser\\
				LCoS\\
				Diffuser}     & \makecell[l]{\cite{Wagadarikar08_CASSI,Wagadarikar09_vCASSI,Gehm07_DDCASSI,Lin14_ACMTG,Cao2011_Cao_Prism,Meng2020_OL_SHEM} \\
				\cite{Yuan15_JSTSP_CASSI_SI}\\
				\cite{Liu_2016_GISR}}\\
			\hline
			$(x,y,t)$ & \makecell[l]{Shifting mask\\
				Streak camera \\ DMD \\ LCoS\\ Structured illumination} & 
			\makecell[l]{\cite{Llull13_OE_CACTI,Koller15_videoCS} \\ \cite{Gao14_Nature_fast} \\ \cite{Hitomi11_ICCV_videoCS,Qiao2020_APLP,Qiao2020_CACTI} \\ \cite{Reddy11_CVPR_P2C2}\\ \cite{Yuan16BOE}}\\
			\hline 
			$(x,y)+z$ & \makecell[l]{Stereo imaging \\ Structured illumination \\
				Depth from focus \\     Focal stack \\Stereo X-ray, lensless}	& \makecell[l]{\cite{Sun17OE} \\ \cite{Sun16OE} \\
				\cite{Llull15Optica} \\  \cite{Llull14_COSI_focalstack} \\ \cite{duarte2019computed}}\\
			\hline
			$(x,y,z)$ & \makecell[l]{Holography \\ Diffuser \\ Coherence tomography \\ X-ray tomography, coded aperture \\
			Light field} & \makecell[l]{\cite{Brady09_ComHolo,Brady18_CPRL_omHolo} \\ \cite{Antipa18_Optica_3D} \\ \cite{Qiao19_OCT,Qiao_CSOCT_arxiv2020}\\ \cite{Brady13_xray,Greenberg:13_xray,Hassan16_xray,MacCabe13_xray,Zhu:18_xray}\\\cite{Babacan12_lightfield,Ruiz13_lightfield,He20_Lightfield_OE}} \\
			\hline
			$(x,y,p)$ & LCoS & \cite{Tsai15OE} \\
			\hline
			$(x,y,d)$ & \makecell[l]{Mask \\ LCoS} & \makecell[l]{\cite{Nayar_2000_CVPR_HDR,Yuan_16_OE} \\ \cite{Nayar_03_ICCV_HDRSLM}} \\
			\hline
			$(x,y,t)+z$  & \makecell[l]{ Stereo imaging\\ Structured illumination  \\ Focal stack}
			& \makecell[l]{ \cite{Sun17OE} \\ \cite{Sun16OE} \\ \cite{Yuan14CVPR}} \\
			\hline
			$(x,y,d,t)$ & Random exposure & \cite{Portz13_ICCP_hdrt} \\
			\hline
			$(x,y,\lambda, t)$ & \makecell[l]{  Shifting mask + disperser\\ LED + DMD}  & \makecell[l]{\cite{Tsai15OL}\\ \cite{Ma2021_LeSTI_OE}} \\
			\hline
			$(x,y,\lambda, p)$ & LCoS & \cite{Tsai13_AO_SpecPol} \\
			\hline
			$(x,y,z,t)$ & DMD + holography & \cite{Wang17_OE_HoloVideo} \\
			\hline
			$(x,y,z,\lambda)$ & X-ray, coded aperture & \cite{Pang14_xray} \\
			\hline
	\end{tabular}}
	\label{Tab:SCI_system}
\end{table}

\subsection{Underlying Principle}
Consider the 3D data-cube shown in the left of Fig.~\ref{Fig:sci_encoder}, which uses a spectral cube 
$(x,y,\lambda)$ as an example. Our  aim is to sample this 3D cube using a 2D detector (sensor). The underlying principle is to compress this 3D cube into a 2D measurement. In contrast to global random matrix CS systems, \eg, the single-pixel camera regime~\cite{Duarte08SPM}, SCI compresses the data-cube across the third dimension, \eg, the spectral dimensional in Fig.~\ref{Fig:sci_encoder}. As depicted in Fig.~\ref{Fig:sci_encoder}, we first decompose the 3D cube into its constituent 2D frames based on different wavelengths. Then, for each 2D frame, we impose a 2D  mask to modulate it via pixel-wise product (element-wise multiplication denoted by $\odot$). These modulated frames are then summed up (integrating light to the sensor within one exposure time) into a coded, and thus {\em compressed}, measurement, which will be 
captured by the 2D detector, \eg, a charge-coupled device (CCD) or a complementary metal–oxide–semiconductor (CMOS) camera. 

This coded measurement is then provided to the algorithm along with the masks to reconstruct the desired 3D data-cube (Fig.~\ref{Fig:sci_decoder}). With these two steps, we earn an optical hardware encoder plus a software decoder SCI system.

The following question is how to devise the hardware to implement the encoding process in Fig.~\ref{Fig:sci_encoder}, probably in a {\em low-cost and low-power} fashion, and {\em efficient and effective} algorithms to reconstruct the desired data-cube (Fig.~\ref{Fig:sci_decoder}). As mentioned in the introduction, various systems have been designed in the literature to capture HD data-cubes. Let $(x,y)$ denote the 2D spatial-dimension, $z$ the depth, $t$ the temporal dimension, $\lambda$ the spectral (or wavelength) dimension, $p$  the polarization and $d$ the dynamic range. Table~\ref{Tab:SCI_system} summarizes the imaging systems designed to capture different data-cubes with references as well as the key modulation techniques. Note that $(x,y,z)$  denotes 3D   tomography and we use $(x,y)+z$ to represent a 2D image plus a depth map, which is also called 2.5D in certain articles in the literature.

\begin{figure}[h!]
	\centering
	{\includegraphics[width=1\columnwidth,height=3.5cm]{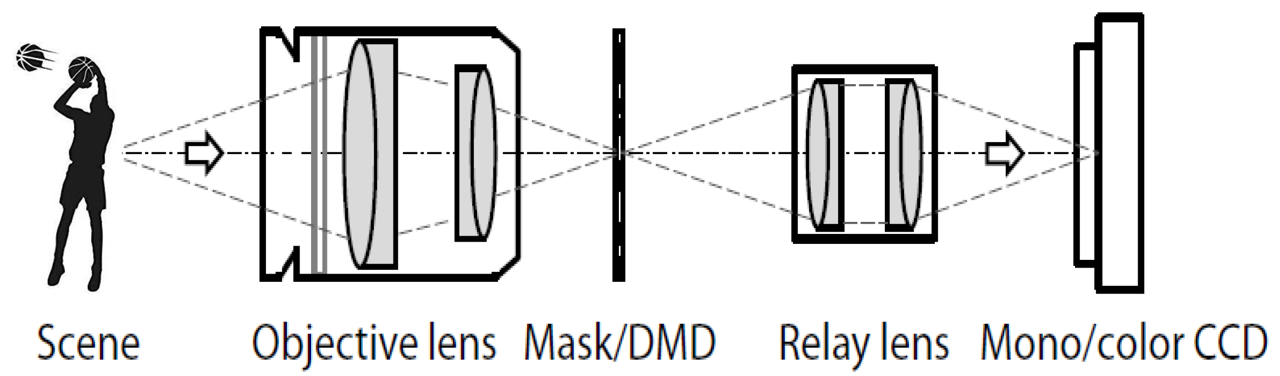}}
	\caption{Schematic of the coded aperture compressive temporal imaging (CACTI) system~\cite{Llull13_OE_CACTI}. A snapshot on the CCD encodes tens of temporal frames of the scene coded by the spatial-variant mask, \eg, the shifting physical mask or different patterns on the digital micromirror device (DMD). The mask/DMD and the mono/color detector, \eg, CCD, are in the conjugate image plane of the scene.}
	\label{Fig:cacti}
\end{figure}

\subsection{Exemplar Hardware SCI Systems \label{Sec:Ex_hard}}
 This section presents two novel designs: one for video SCI and the other one for spectral SCI. 
In video SCI using a low-speed camera to capture high-speed scenes, we use the coded aperture compressive temporal imaging (CACTI)~\cite{Llull13_OE_CACTI} as an example shown in Fig.~\ref{Fig:cacti}. The high-speed scene is collected by the objective lens and spatially coded by the temporal-variant mask, such as the shifting mask or different patterns on the digital micromirror device (DMD) or the spatial light modulator (SLM), \eg, the liquid crystal on silicon (LCoS). Then the coded scene is detected by a CCD. A snapshot on the CCD encodes tens of temporal frames of the high-speed scene. The number of coded frames for a snapshot is determined by the number of variant codes of the mask or different patterns on the DMD within the integration (exposure) time.
It is easy to connect the video SCI setup with the encoding process of SCI in Fig.~\ref{Fig:sci_encoder} as the different channels are now high-speed frames at different timestamps and the masks are the shifting variants of the physical mask or the coding patterns displayed on the DMD.

\begin{figure}[h!]
	\centering
	{\includegraphics[width=1\columnwidth]{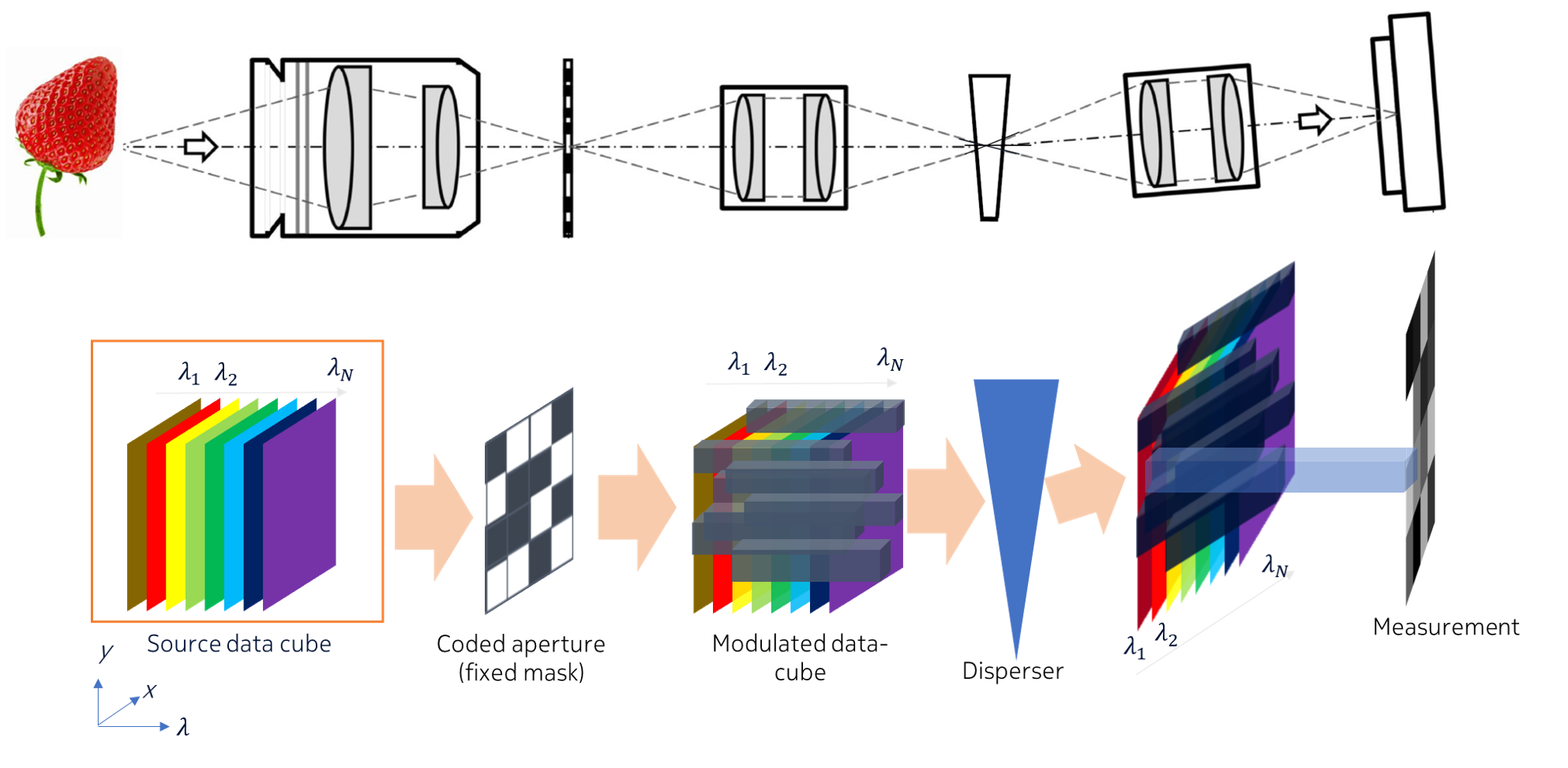}}
	\caption{Schematic (top row) and sampling principle (bottom row) of the coded aperture snapshot spectral imaging (CASSI) system~\cite{Wagadarikar08_CASSI}. A snapshot on the CCD encodes tens of spectral bands of the scene spatially coded by the mask and spectrally coded by the dispersive element (prism or grating). The mask, the dispersive element and the CCD are in the conjugate image plane of the scene. In the bottom part, we used a $4\times 4$ checkerboard to demonstrate mask but in real system, it is on the order of thousands by thousands elements; the `black' cell blocks the light and the `white' one transmits the light.}
	\label{Fig:cassi}
\end{figure}

For the spectral SCI, we use the single disperser coded aperture compressive spectral imager (CASSI)~\cite{Wagadarikar08_CASSI} shown in Fig.~\ref{Fig:cassi}. In CASSI, the spectral scene, \ie, the $(x,y,\lambda)$ data-cube, is collected by the objective lens and spatially coded by a {\em fixed} mask, and then the coded scene is spectrally dispersed by the dispersive element, such as a prism or a grating. The spatial-spectral coded scene is detected by the CCD. A snapshot on the CCD encodes tens of spectral bands of the scene.

While it is straightforward to understand the video SCI encoding process, it is not intuitive to connect the coding process in CASSI with Fig.~\ref{Fig:sci_encoder}.
We demystify the simple and efficient coding process of CASSI here. The main idea is using  a {\em fixed mask plus a disperser} to implement multiple ($N$ in Fig.~\ref{Fig:sci_encoder}) different masks and then to modulate different spectral channels.
 
Let $\Fmat \in {\mathbb R}^{N_x \times N_y \times N_{\lambda}}$ denote the 3D spatial-spectral cube shown in the bottom-left of Fig.~\ref{Fig:cassi} and $\Mmat^0 \in  {\mathbb R}^{N_x \times N_y}$ the (fixed) physical mask used for signal modulation.
We use ${\Fmat}' \in {\mathbb R}^{N_x \times N_y \times N_{\lambda}}$ to represent the modulated signals where images at different wavelengths are modulated separately (but by the same mask now), \ie, 
for $n_{\lambda} = 1,\dots, N_{\lambda}$, we have
\begin{equation}
{\Fmat}' (:,:,n_{\lambda}) = {\Fmat} (:,:,n_{\lambda}) \odot \Mmat^0,
\end{equation} 
where $\odot$ represents the element-wise multiplication.
Recalling the encoding process of SCI in  Fig.~\ref{Fig:sci_encoder}, we need each spectral channel to be modulated by a {\em different} mask. However, we have until now that all spectral channels are modulated by the {\em same} mask and we have not achieved yet what we expect during encoding. 

{Next comes the disperser, whose role is to disperse the light to different spatial locations based on their wavelengths}. After this modulated cube (by the same mask) passes the disperser, $\Fmat'$ is {\em tilted} and is considered to be sheared along the $y$-axis. 
We then use $\Fmat''\in {\mathbb R}^{N_x \times (Ny + N_{\lambda}-1) \times N_{\lambda}}$ to denote the tilted cube 
and assuming $\lambda_c$ to be the reference wavelength, \ie, image $\Fmat'(:,:,n_{\lambda_c})$ is not sheared along the $y$-axis,
%
we thus arrive at
\begin{equation}
\Fmat'' (u,v, n_{\lambda}) = \Fmat'(x, y + d(\lambda_n - \lambda_c), n_{\lambda}),
\end{equation}
where $(u,v)$ indicates the coordinate system on the detector plane, and $\lambda_n$ is the wavelength at the $n_\lambda$-th channel. 
Here, $d(\lambda_n -\lambda_c)$ signifies the spatial shifting  for the $n_\lambda$-{th} channel.
The compressed measurement at the detector $y(u,v)$ can thus be modeled as 
\begin{equation}
{\textstyle y(u,v) = \int_{\lambda_{\rm min}}^{\lambda_{\rm max}} f'' (u,v, n_{\lambda}) d \lambda,}
\end{equation}
since the sensor integrates all the light in the wavelength range $[\lambda_{\rm min}, \lambda_{\rm max}]$, where $f''$ is the analog (continuous) representation of $\Fmat''$.
In discretized form, the captured 2D measurement $\Ymat \in {\mathbb R}^{N_x \times (Ny + N_{\lambda}-1)}$ is 
\begin{equation}
{ \Ymat = \sum_{n_{\lambda}=1}^{N_{\lambda}}  \Fmat'' (:,:, n_{\lambda}) + \Emat},
\end{equation}
which is a {\em compressed} frame containing the information of all modulated spectral channels and $\Emat\in {\mathbb R}^{N_x \times (Ny + N_{\lambda}-1)}$ represents the measurement noise.

For convenience of the model description, we further set 
$\Mmat\in \mathbb{R}^{N_x \times (Ny + N_{\lambda}-1) \times N_{\lambda}}$ to be the {\em shifted version} of the (same physical) mask corresponding to different wavelengths, \ie,
\begin{equation}
\Mmat (u,v, n_{\lambda}) = \Mmat^0(x, y + d(\lambda_n - \lambda_c)).
\end{equation}
Similarly, for each signal frame at different wavelengths, the shifted version $\tilde{\Fmat} \in {\mathbb R}^{N_x \times (Ny + N_{\lambda}-1) \times N_{\lambda}}$ is given by
\begin{eqnarray}
\tilde{\Fmat} (u,v, n_{\lambda}) = \Fmat(x, y + d(\lambda_n - \lambda_c), n_{\lambda}). \label{Eq:Fmat_Ft}
\end{eqnarray}
Based on the above, the measurement $\Ymat$ can be represented as
\begin{equation}
{ \Ymat = \sum_{n_{\lambda}=1}^{N_{\lambda}}   \tilde{\Fmat} (:,:, n_{\lambda})  \odot \Mmat (:,:, n_{\lambda}) + \Emat}. \label{Eq:sensing_Matrix}
\end{equation}
Eq.~\eqref{Eq:sensing_Matrix} corresponds to the encoding process of SCI in Fig.~\ref{Fig:sci_encoder} and also leads to the similar forward model used in Sec.~\ref{Sec:Model}.

Note that the 3D mask $\Mmat$ can be obtained by calibration and after we solve for $\tilde{\Fmat}$ given $\Ymat$ and $\Mmat$, we can obtain the desired 3D cube by shifting it back to $\Fmat$ based on the relationship in \eqref{Eq:Fmat_Ft}. 


\subsection{Other Modulation Methods for HD Signals}
Now we briefly introduce the modulation methods in other SCI systems.
While a physical mask and a DMD can only be used for amplitude modulation, 
an LCoS can implement amplitude, phase, polarization and spectral modulation. Therefore, a single LCoS can perform spectral SCI by replacing the mask plus disperser~\cite{Yuan15_JSTSP_CASSI_SI}, though the spectral resolution will depend on the response of the LCoS. LCoS has thus also been used in polarization, depth and dynamic range SCI.
The reason that LCoS is not usually used in video SCI is due to the slow refresh rate. 
Another amplitude modulation method is structured illumination, which needs an additional light source to illuminate the scene. This leads to an active system and thus the power consumption might be larger than that of the other modulation methods. 
Another advantage of structured illumination method is that the illumination patterns can have different scales at different depths, and hence they can resolve the depth information of the scene~\cite{Sun16OE}.
In addition to this method to resolve depth, a liquid lens changing its focal depth in one exposure time~\cite{Yuan14CVPR}, and thus actively changing the blur kernels~\cite{Llull15Optica}, and stereo vision~\cite{Sun17OE} have also been used for depth SCI. 

In another line of research the spatial/spectral/depth responses of the object to some medium, like diffuser~\cite{Antipa18_Optica_3D} or ground glasses are used to achieve the modulation and thus to implement SCI.

\subsection{Summary \label{Sec:hard_summary}}
CASSI and CACTI are two typical SCI systems, with CASSI representing passive modulation and CACTI active modulation.
From a power consumption perspective, passive modulation is preferred and thus CASSI is a favorite design for SCI.

From an optical perspective, single-shot compressive imaging systems typically have two coding regimes: pupil coding and image-space coding.  Pupil coding is adopted in lensless cameras and diffraction tomography (holograpy and radar) systems. In holographic systems pupil coding may consist of a simple measurement of the Fourier transform, but modulation can also be incorporated in either coherent or incoherent pupil coded systems.  Instead of using an imaging lens, a random medium, which can be a scattering medium, a mask or a random reflective surface, is placed in front of the detector. Each spatial point in the object space generates a different random pattern on the detector. In the reconstruction, a 2D sensing matrix is constructed by vectorizing the random patterns corresponding to each object point as each column of the matrix. Therefore, the matrix size and thus the computation complexity scales with the square of the measurement image height (assuming a square image). By contrast, in the image-space coding regime, two cascaded relay lenses are typically employed. The first lens relays the object to the coded aperture plane and the second lens relays the coded object to the detector plane. A modulator is placed on the coded aperture plane, or between the aperture and the detector, or on the detector plane, to shear the coded object cube before it collapses to a 2D measurement on the detector. In this regime, the sensing matrix in the reconstruction does not necessarily have to be an expanded 2D matrix. Instead, it can be a 3D matrix with each page (each transverse plane) of the matrix being a shifted version of the aperture pattern. Therefore, the matrix size scales only with the measurement image height. 
In this article, we mainly consider the image-space coding for SCI and thus the calibration is easier. This lead to the forward model described in the next section.

\section{Mathematical Model of SCI \label{Sec:Model}}
\begin{figure}[h!]
	\centering
	{\includegraphics[width=1\columnwidth]{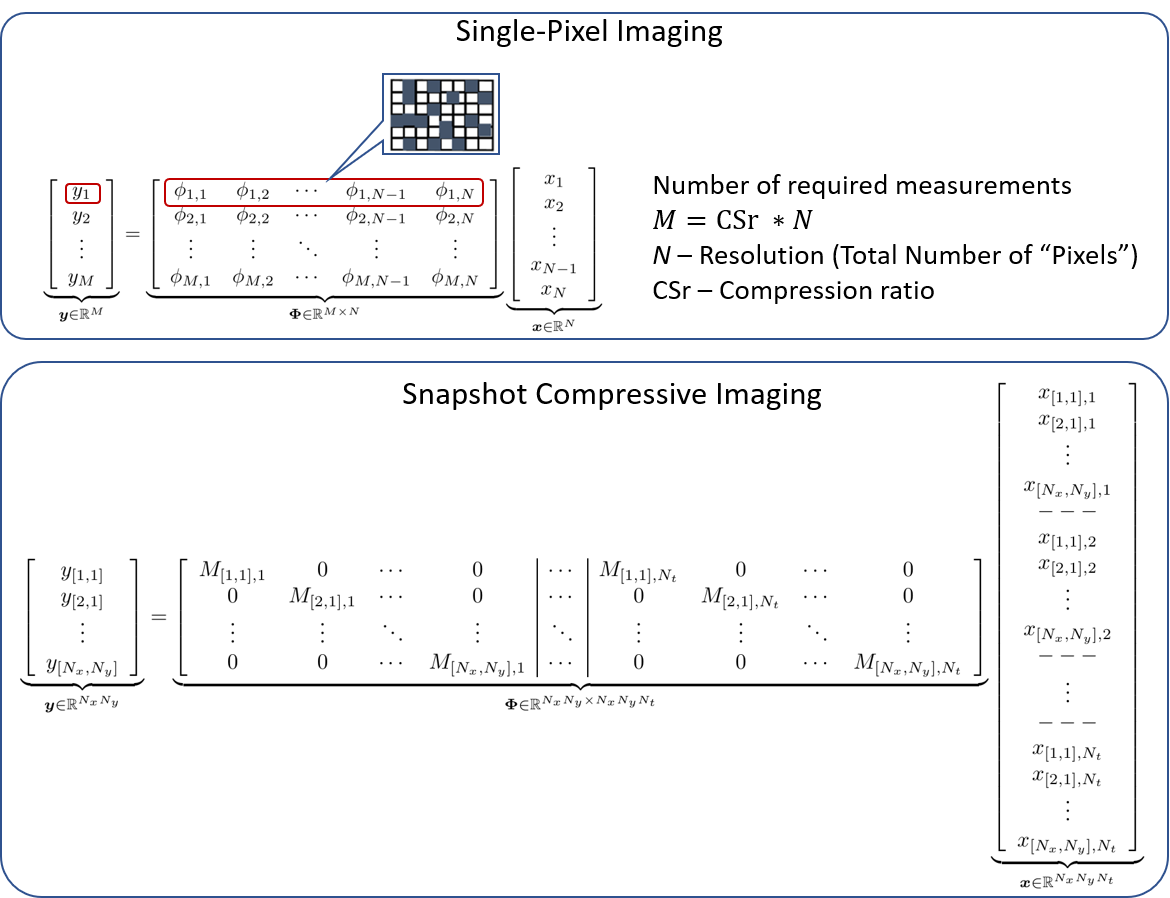}}
	\caption{Comparing the forward model of single-pixel imaging (top)~\cite{Duarte08SPM} and snapshot compressive imaging (bottom). The signal-pixel imaging system usually captures a still 2D image, denoted by the vector $\xv \in {\mathbb R}^N$. Each measurement (each element in $\yv\in {\mathbb R}^M$)  is captured by imposing a modulation pattern (each row in the sensing matrix $\Phimat \in {\mathbb R}^{M\times N}$) on the 2D image.  In an SCI system, the measurement is a 2D image $\Ymat \in {\mathbb R}^{N_x\times  N_y}$. Each frame in the 3D signal $\Xmat \in {\mathbb R}^{N_x \times N_y \times N_t}$ is modulated by a different mask resulting in the sensing matrix $\Phimat \in {\mathbb R}^{N_xN_y\times N_xN_yN_t}$. Each element in $\Ymat$ is a weighted sum of the corresponding elements in each frame of $\Xmat$. Obviously, the compressive sensing ratio of SCI is $1/N_t$ (in single-pixel imaging, it is $M/N$). }
	\label{Fig:forward}
\end{figure}

In this section, we employ a unified mathematical model for both video and spectral (and any other 3D) SCI. Without considering optical details, let $\Xmat \in {\mathbb R}^{N_x\times N_y\times N_t}$ denote the 3D video data-cube. The mask  $\Mmat\in {\mathbb R}^{N_x\times N_y\times N_t}$ is employed to modulate the data-cube. Let $\Xmat_k \in {\mathbb R}^{N_x\times N_y}, \forall k=1,\dots,N_t$, denote the  $k$-th frame of the data-cube and similarly  $\Mmat_k$  the corresponding  $k$-th mask. The 2D measurement $\Ymat$ can be modeled as
\begin{equation}
\Ymat = \sum_{k=1}^{N_t} \Xmat_k \odot \Mmat_k + \Emat, \label{Eq:Y_sum}
\end{equation}
where $\Emat$ again denotes the measurement noise as in Eq.~\eqref{Eq:sensing_Matrix}. Define
\begin{eqnarray}
\xv &=& [\xv_1\ts, \dots, \xv_{N_t}\ts]\ts,  \label{Eq:x1toNt}\\
\Phimat &=& [\Dmat_1, \dots, \Dmat_{N_t}], \label{Eq:Phimat}
\end{eqnarray}
where $\xv_k = {\rm vec}({\Xmat_k})$ represents the vectorization of the $k$-th frame (by stacking columns) and $\Dmat_k = {\rm Diag}({\rm vec}(\Mmat_k))$ is a diagonal matrix with diagonal elements the vectorized form of $\Mmat_k$. We obtain the following forward model
\begin{equation}
\yv = \Phimat \xv + \ev, \label{Eq:yPhix}
\end{equation}
where $\yv = {\rm vec}(\Ymat)$ and $\ev = {\rm vec}(\Emat)$. This formulation is similar to CS, but with a sensing matrix $\Phimat$ having a special structure shown in the right-part of Fig.~\ref{Fig:forward}.
Consequently, the {\em sampling rate} here is equal to  $1/N_t$. 
The CASSI model is a little bit more complicated as derived in Sec.~\ref{Sec:Ex_hard} but we can still arrive at the representation of \eqref{Eq:yPhix} by replacing some parameters.

The main difference between SCI and single-pixel imaging~\cite{Duarte08SPM} lies in the forward model. Specifically, 
\begin{itemize}
	\item In the single-pixel camera, the sensing matrix is a dense matrix.  Each row of $\Phimat$ corresponds to one pattern of the modulator imposed on the scene (a 2D still image) $\xv$ and the single-pixel detector captures one measurement (one element in $\yv$).
	\item In SCI, the sensing matrix $\Phimat$ is a sparse matrix, which is a concatenation of $N_t$ diagonal matrices. Each element of the measurement is a weighted summation of the corresponding elements in $\xv$ (a 3D or HD cube) across $N_t$ frames, modulated by the masks.
\end{itemize}
This difference (depicted in Fig.~\ref{Fig:forward}) will also affect the design of the reconstruction algorithm for these two imaging systems. In single-pixel imaging, $\Phimat$ is a large matrix and if random, we need to save it.
An alternative way is to use a structural matrix (\eg, DCT or Hadamard matrix~\cite{Gopalsami12_Hadamard,Spinoulas:12_hadamard}) with some permutations~\cite{Yuan2020TMM} or some designed matrices~\cite{Tsiligianni14_TIT}. This will help the computation since fast transformations can be used. By contrast, in SCI, even though $\Phimat$ is bigger than the one in single-pixel imaging, we do not need to store the $\Phimat$ matrix but only the masks (the $N_t$  diagonal elements in $\Phimat$). Furthermore, due to the special structure of $\Phimat$, we observe that $\Phimat\Phimat\ts$ is a diagonal matrix and this has been used in the literature to speed up the reconstruction algorithm for SCI~\cite{Liu19_PAMI_DeSCI}. 

In the following sections, we describe the recently developed theoretical guarantees for SCI reconstruction by considering the special structure of the sensing matrix followed by diverse algorithm designs.

\section{Theoretical Guarantees of SCI \label{Sec:Theory}}
Although the SCI imaging systems have been built for more than a decade (counting the first CASSI system in 2007~\cite{Gehm07_DDCASSI}), only recently solid theoretical results have been developed~\cite{Jalali19TIT_SCI}, which take into account the special structure of the sensing matrix in Eq.~\eqref{Eq:Phimat}.
{The theoretical derivation is based on signal compression results applied to compressive sensing~\cite{jalali2016compression}. Before presenting the main result, we provide some necessary definitions.}

\subsection{Data Compression for SCI Signals}
{Consider a compact set ${\cal Q}\subset\mathbb{R}^{N_xN_yN_t}$. As defined in \eqref{Eq:x1toNt}, each signal $\xv\in{\cal Q}$
consists of $N_t$ frames $\{\xv_1,\dots,\xv_{N_t}\}$ in $\mathbb{R}^{N_xN_y}$.  A lossy compression {\em code} of rate $r\in\mathbb{R}^+$ for ${\cal Q}$  is characterized by its {\em encoding mapping} $f$, where
\begin{align}
{ f:{\cal Q} \rightarrow \left\{1,2,\ldots, 2^{N_xN_yN_t r}\right\},} \label{eq:f-def}
\end{align}
and its {\em decoding mapping} $g$, where
\begin{align}
{ g:\left\{1,2,...,2^{N_xN_yN_t r}\right\}\gos \mathbb{R}^{N_xN_yN_t}.}\label{eq:g-def}
\end{align}
The {\emph {average  distortion}} between  $\xv$ and its reconstruction  $\hat{\xv}$ is defined as 
\begin{equation}
{ d(\xv,\hat{\xv})\triangleq   {1\over  {N_xN_yN_t}}\|\xv-\hat{\xv}\|_{2}^{2}\;}.\label{eq:def-delta-B}
\end{equation}
Let $\tilde{\xv} = g(f(\xv)).$
The  distortion  of the {\em code} $(f,g)$ is denoted by $\delta$, which is defined as the supremum of all achievable distortions. That is,
\begin{equation}
{ \delta\triangleq \sup_{\xv\in\mathcal{Q}} d(\xv,\tilde{\xv}) = \sup_{\xv\in\mathcal{Q}} {1\over {N_xN_yN_t}}\|\xv-\tilde{\xv}\|_{2}^{2}.}
\end{equation}
Let  ${\cal C}$ denote the {\em codebook} of this code defined as
\begin{equation}
{ {\cal C}=\{g(f(\xv)):\; \xv\in\cal{Q}\}.}  \label{eq:codebook}
\end{equation}
Clearly, since the code is of rate $r$, $|{\cal C}|\leq 2^{N_xN_yN_tr}$, where $|{\cal C}|$ denotes the cardinality of $\cal C$. 
Consider  a family of compression codes $\{(f_r,g_r)\}_r$ for set  ${\cal Q}$,  indexed by their rate $r$. The deterministic {\em distortion-rate function} for this family of codes is defined as 
\begin{equation}
\delta(r)=\sup_{\xv\in{\cal Q}} {1\over{N_xN_yN_t}}\|\xv-g_r(f_r(\xv))\|_2^2.
\end{equation}
The corresponding deterministic {\em rate-distortion function} for this family of codes is defined as
\begin{equation}
    r(\delta)=\inf\{r: \delta(r)\leq \delta\}.
\end{equation}}
{In the theoretical derivation, we assume that the compression code $g(f(\xv))$ return a codeword in $\cal C$ that is closest to $\xv$. This is,
\begin{equation}
    g(f(\xv))=\argmin_{\cv\in{\cal C}}\|\xv-\cv\|_2^2. \label{Eq:gfx}
\end{equation}}
Note that there are no guarantees regarding the computational feasibility of \eqref{Eq:gfx}.

\subsection{Theoretical Guarantee by Compression-based Recovery}
A compressible signal pursuit (CSP)-type optimization was proposed in~\cite{Jalali19TIT_SCI} as a compression-based recovery algorithm for SCI.
Consider the compact set ${\cal Q}\subset\mathbb{R}^{N_xN_yN_t}$ equipped with a rate-$r$ compression code described by mappings  $(f,g)$, defined in \eqref{eq:f-def}-\eqref{eq:g-def}. 
Consider $\xv\in\cal{Q}$ and the reconstructed signal $\hat\xv$ is obtained by solving the following optimization:
\begin{equation}
    \hat\xv = \argmin_{\cv \in {\cal C}} \|\yv - \Phimat \cv\|_2^2, \label{eq:CSP}
\end{equation}
where ${\cal C}$ is defined in \eqref{eq:codebook} and $\Phimat$ is the sparse acquisition matrix.
In other words, given a measurement vector $\yv$, this optimization, among all compressible signals, \ie, signals in the codebook, picks the one that is closest to the observed measurement when sampled according to $\Phimat$. Note again that, similar to \eqref{Eq:gfx}, there are no guarantees regarding the computational feasibility of \eqref{eq:CSP}.

The following theorem characterizes the performance of SCI recovery using CSP-type optimization  by connecting the parameters of the (compression/decompression) code, its rate $r$ and the corresponding distortion $\delta$, to the number of frames $N_t$, which determines the compressive sampling ratio and the resulting overall reconstruction quality. 

\begin{theorem}\label{thm:main-csp}
	\cite{Jalali19TIT_SCI} Assume that $\forall \xv\in{\cal Q}$, $\|\xv\|_{\infty}\leq {\frac{\rho}{2}}$. Further assume  the rate-$r$ code achieves distortion $\delta$ on $\cal{Q}$.
	Moreover, assuming each element in $\Mmat$ is drawn from the standard Gaussian distribution, \ie, $M_{i,j,k}\stackrel{\rm i.i.d.}{\sim} {\cal N}(0,1), \forall i=1,\dots, N_x; j=1, \dots, N_y; k=1,\dots, N_t $. Let $\hat{\xv}$ denote the solution of compressible signal pursuit optimization in \eqref{eq:CSP}.  Assume that $\epsilon>0$ is a free parameter, such that  $\epsilon \leq {\frac{16}{3}}$. Then,
	\begin{align}
	{ {\frac{1}{N_x N_y}}\|\xv-\hat{\xv}\|_2^2\leq N_t (\delta +\rho^2\epsilon)}, \label{Eq:them_error}
	\end{align}
	with a probability larger than $ { 1- 2^{N_xN_yN_tr+1} {\rm e}^{- N_xN_y(\frac{3\epsilon}{32})^2}}$.
	%
\end{theorem}
Note that we assume that the signal is bounded, \ie, $\|\xv\|_{\infty}\leq {\frac{\rho}{2}}$, which on one hand is necessary for the proof of the theorem, while on the other hand, image and video pixel values are usually bounded after captured by a camera (through the dynamic range of the sensor).
Theorem~\ref{thm:main-csp} tells us that given a compressible signal parameterized by the compression rate $r$ and distortion $\delta$, if this signal is compressively captured by an SCI system, the reconstruction error (the error between the estimated and ground truth signals) is bounded with high probability by the distortion. 
In the following, we provide some observations resulting from the theorem.
\begin{itemize}
    \item Consider that the spatial size of the signal is fixed, \ie, $N_x$ and $N_y$ are fixed, and further assume that the distortion of the signal, $\delta$, is fixed. Given a pre-defined $\rho$ (by the camera) and $\epsilon$, from \eqref{Eq:them_error}, a larger $N_t$ will lead to a larger reconstruction error. 
    This is intuitive since the more video frames we want to reconstruct, the more challenging the reconstruction algorithm.
    \item Theorem~\ref{thm:main-csp} is based on the discrete quantized space of the compressed signal (right hand side of~\eqref{eq:f-def}). There still needs some research to extend the result to the continuous space.
    \item The proof of Theorem~\ref{thm:main-csp} (details in~\cite{Jalali19TIT_SCI}) essentially includes two steps: a) the determination of an error bound for a fixed value of the signal $\xv$ in the encoded space based on the concentration of measure, and b) the determination of a union bound to collect every possible $\xv$ in the encoded space.  
    In particular, $\epsilon$ is a parameter of the concentration of measure determining how long the tail of a distribution being covered in the bound is. If $\epsilon$ is larger, we can cover a longer tail, and the probability of the success rate  will be higher, but the error bound will be larger, too.
    Specifically, on the right hand side of~\eqref{Eq:them_error}, if $\epsilon$ is larger, the reconstruction error will be larger and also the probability $ { 1- 2^{N_xN_yN_tr+1} {\rm e}^{- N_xN_y(\frac{3\epsilon}{32})^2}}$ will be larger.
    \item From the system design perspective, we would like to see the direct relationship between the sampling rate $1/N_t$ and the reconstruction error. This is connected by the rate-distortion function of the signal, \ie, the $(r, \delta)$ pair. In~\cite{Jalali19TIT_SCI}, a corollary is derived to make this point clear.
    \begin{corollary}\label{corr:main-csp}
    	\cite{Jalali19TIT_SCI} Consider the same setup as in Theorem~\ref{thm:main-csp}. Given $\eta>0$, assume that
    	\begin{equation}
    	    N_t < \frac{1}{\eta} \left(\frac{\log {1\over\delta}}{2r}\right). \label{Eq:N_t}
    	\end{equation}
    	Then,
    	\begin{equation}
    	    {\rm Pr}\left(\frac{1}{N_xN_yN_t}\|\xv - \hat{\xv}\|_2^2 > \delta + 8\rho^2 \sqrt{\frac{\log{1\over \delta}}{\eta}}\right) \le 2 e^{-\frac{\log{1\over\delta}}{5\eta}N_xN_y}, \label{Eq:Pr_corr}
    	\end{equation}
    	where ${\rm Pr} (~)$ denotes probability. 
     \end{corollary}
     The corollary is obtained by setting $\epsilon = 8 \sqrt{\frac{\log{1\over \delta}}{\eta}}$ in Theorem~\ref{thm:main-csp}. In this way, $N_t$ is removed from the probability bound on the right hand side of \eqref{Eq:Pr_corr}. By setting $N_t$ according to~\eqref{Eq:N_t}, both the reconstruction error and the probability are bounded by the distortion $\delta$, given $N_x, N_y, \rho$ and $\eta$.   
     Clearly, the smaller the distortion $\delta$ due to coding the larger the $N_t$, \ie, more frames can be compressed into a single measurement, and the smaller the reconstruction error with high confidence. 
\end{itemize}

Note that the result from Theorem~\ref{thm:main-csp} can be applied into any SCI systems with the forward model shown on the lower part of Fig.~\ref{Fig:forward}. 
The key assumption is that the desired HD data is highly compressible, which is in general true for the designed SCI systems. 
Essentially, as mentioned before, since the compressive sampling rate of SCI is equal to  $1/N_t$, the reconstruction error is basically bounded by the inherent compressibility of the signal. 

\subsection{Limitations and Discussions of The Theoretical Results}

{This theoretical finding provides us with some insights into the SCI systems and also some guidance on the algorithm design. However, Theorem~\ref{thm:main-csp} shares the limitations of other CS based theoretical analysis.
In the following, we first list the limitations:}

\begin{itemize}
    \item {The theoretical result is based on CSP, which is not computational feasible. As mentioned in~\cite{Jalali19TIT_SCI}, finding the solution of the CSP optimization through exhaustive search over the codebook is infeasible, even for small values of the video size}. 
    \item It is hard to verify the values of the parameters (such as $r, \delta$) in real applications.
	\item The mask is considered to be i.i.d. Gaussian, which is unrealistic in real SCI systems. Specifically, the mask or DMD response of light is always non-negative. Therefore, there is a gap, or at least a constant---probably a DC (direct current) term---between this theorem and real SCI systems. {This gap arises from the `zero-mean' requirement in the proof of the theorem.}
	\item During the implementation of modulation, especially using a DMD, a binary mask $\{0,1\}$ is generally used.  
	\begin{itemize}
		\item In this case, an extended theory by considering the mask being the Rademacher distribution (\ie, random variables taking values 1 or -1 with equal probability) will be closer to real applications. An explicit bound (similar to Eq.~\eqref{Eq:them_error}) can be derived using the concentration of measure as used in~\cite{Wang15_SIAM} {by considering the Rademacher distribution for the mask}.
		\item Another solution stems from the hardware side. By employing a beamsplitter into the optical path~\cite{Yuan15_JSTSP_CASSI_SI,Yuan17AO} and imposing two conjugate modulation patterns (one with $\{0,1\}$ and the other one with $\{1,0\}$, when one cell of the first mask is 1, the corresponding cell of the second mask is 0) and then using the subtraction of these two measurements, we can get the measurement derived in the theorem.   
		This solution can also potentially increase the dynamic range of the measurement and thus improve the reconstruction quality as discussed in Sec.~\ref{Sec:dynamicRange}.
	\end{itemize} 
\end{itemize}

\begin{figure}[h!]
	\centering
	{\includegraphics[width=1\columnwidth]{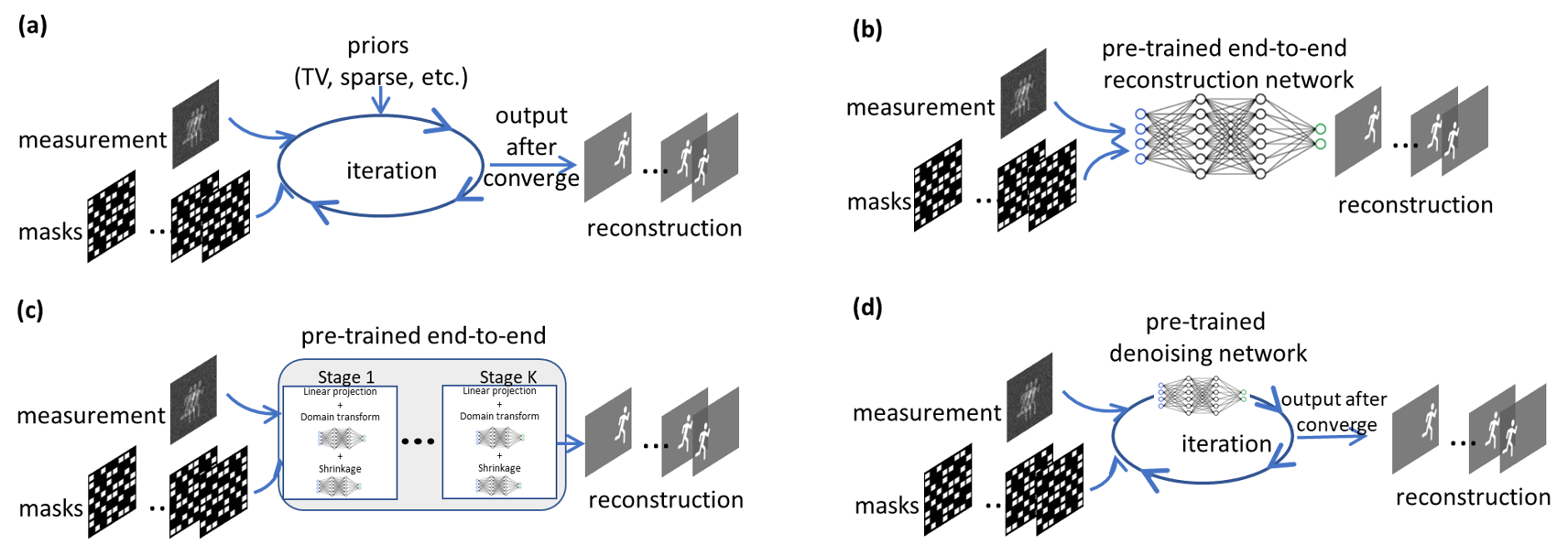}}
	\caption{Different frameworks of the reconstruction algorithm for SCI: (a) Conventional optimization based iterative algorithm. (b) End-to-end deep learning based on convolutional neural networks, (c) deep unrolling/unfolding algorithms, where $K$ small CNNs are used and (d) Plug-and-Play algorithms using pre-trained denoising network as priors.}
	\label{Fig:DL}
\end{figure}

It is also important to discuss the guidance provided by this theoretical finding for the algorithmic design. 
\begin{itemize}
    \item As mentioned above, directly finding the solution of CSP is infeasible. However, as developed in \cite{Jalali19TIT_SCI}, iterative algorithms such as compression-based projection gradient descent can be developed to approximate the solution. 
    In this case, the denoising step such as \eqref{Eq:denoise} in the algorithm can be recognized as a code (encoding-decoding mappings) to play the role of \eqref{Eq:gfx}.   
    \item Assuming the above solution is a good approximation to CSP, then the main reconstruction error term in~\eqref{Eq:them_error} is bounded by the distortion of the code, \ie, $\delta$. Therefore, a better code (with a smaller $\delta$) of the SCI signal can potentially achieve a better result. This is consistent with the experimental results in Sec.~\ref{Sec:Real} and has also been observed in the literature~\cite{Yuan2020_CVPR_PnP}.
    \item {If we treat the deep denoiser (the deep denoising network) as a code, and since the performance of deep denoiser is usually higher than conventional denoising algorithms, it is expected that the plug-and-play algorithm using a deep denoiser (described in Sec.~\ref{Sec:PnP} and shown in  Fig.~\ref{Fig:DL}(d)) leads to good results for SCI reconstruction. This is consistent with the theoretical analysis in Theorem~\ref{thm:main-csp}.}
    \item {However, the end-to-end deep learning algorithms described in Sec.~\ref{Sec:Deep} (shown in Fig.~\ref{Fig:DL}(b)) might not be able to perform the analysis based on this theoretical result, since it is challenging to define the encoder and decoder described in the theorem.}
\end{itemize}


{Bearing the above limitations and guidance of the theorem in mind, the next question of SCI is to solve the inversion problem}, \ie, to reconstruct the desired HD data-cube from the compressed measurement and the masks (Fig.~\ref{Fig:sci_decoder}). 
{Note that the CSP-type optimization in the theorem is not computational feasible and thus other types of algorithms need to be developed.}
During the past decade, various algorithms have been employed and developed. Fig.~\ref{Fig:DL} demonstrates the conventional iteration-based algorithms in (a), and three deep-learning based frameworks in (b-d). Table~\ref{Tab:SCI_algo} summarizes the algorithms being used/developed in various SCI systems.

\begin{table}[!htbp]
	\caption{Different algorithms for various SCI systems. TV: total variation, GMM: Gaussian mixture model.}
	\centering
	{
		\begin{tabular}{|l|l|l|}
			\hline
			Method& Algorithms & SCI Systems \\
			\hline
		Sparse priors & \makecell[l]{GPSR~\cite{Figueiredo07GPSR}\\
				Wavelet-based regularization~\cite{Reddy11_CVPR_P2C2,Yuan14CVPR}\\
				Group sparsity-based regularization~\cite{Wang17_PAMI_GSC}}     & \makecell[l]{$(x,y,\lambda)$ \\
				$(x,y,t)$ \\
				$(x,y,\lambda)$ }\\
			\hline
		     TV priors& \makecell[l]{TwIST~\cite{Bioucas-Dias2007TwIST}\\
				GAP-TV~\cite{Yuan16ICIP_GAP} } & 
			\makecell[l]{$(x,y,\lambda)$, $(x,y,t)$, $(x,y,t,\lambda)$ \\ $(x,y,p)$, $(x,y,t)+z$, $(x,y,z)$}\\
			\hline 
			GMM & \makecell[l]{Off-line training~\cite{Yang14GMM}\\ On-line learning~\cite{Yang14GMMonline}}	&\makecell[l]{$(x,y,t)$ \\ $(x,y,\lambda)$, $(x,y,t)$ }\\
			\hline
		     Dictionary learning & \makecell[l]{3D K-SVD~\cite{Hitomi11_ICCV_videoCS} \\ Bayesian~\cite{Yuan15_JSTSP_CASSI_SI}} & \makecell[l]{$(x,y,t)$ \\ $(x,y,\lambda)$} \\
			\hline
			Deep learning & \makecell[l]{End-to-End~\cite{Iliadis18DSPvideoCS,Miao19ICCV,Iliadis_DSP2020,Qiao2020_APLP,Xiong_17_CVPRW_HSCNN,Meng2020_OL_SHEM,Cheng20ECCV_Birnat,Meng20ECCV_TSAnet} \\ Deep Unrolling~\cite{Wang19_CVPR_HSSP,Ma19ICCV} \\ Plug-and-Play~\cite{Qiao2020_APLP,Qiao2020_CACTI,Yuan2020_CVPR_PnP} }&  \makecell[l]{$(x,y,t)$, $(x,y,\lambda)$ \\ $(x,y,t)$, $(x,y,\lambda)$ \\ $(x,y,t)$}\\
			\hline
	\end{tabular}}
	\label{Tab:SCI_algo}
\end{table}

\section{Regularization Based Optimization Algorithm \label{Eq:regu_opt}}
To solve the ill-posed problem in \eqref{Eq:yPhix}, a regularization term $R(\xv)$, a.k.a., a prior, is usually employed in conventional optimization algorithms.
{The goal of this term is to confine the solution to the desired signal space.} 
These algorithms aim to find an estimate $\hat{\xv}$ of $\xv$ by solving the following problem 
\begin{equation}
{\hat \xv} = \argmin_{\xv} \frac{1}{2}\|\yv - \Phimat \xv\|_2^2 + \lambda R(\xv), \label{Eq:reg}
\end{equation}
where $\lambda$ is a parameter to balance the fidelity term ($\|\yv - \Phimat \xv\|_2^2$) and the regularization term. Eq.~\eqref{Eq:reg} is usually solved by iterative algorithms and various solvers exist based on different forms of $R(\xv)$~\cite{Lucas18SPM}. In SCI, different priors have been used, including sparsity, \eg, the coefficients of the data-cube being sparse in some transform domain~\cite{Yang20_TIP_SeSCI}, and total variation (TV)~\cite{Yuan_TVSCI_arxiv2020}. TV is typically efficient and it usually leads to decent results in real data. Most recently, some complicated priors, \eg, the nonlocal low-rank prior of patch groups have also been used. The DeSCI algorithm, developed in~\cite{Liu19_PAMI_DeSCI} has led to state-of-the-art results (among optimization-based algorithms) for SCI. Regarding the solver, the alternating direction method of multipliers (ADMM)~\cite{Boyd11ADMM} has become popular and rather straightforward to adapt to different systems and leads to good results. 

It has been pointed out in several papers~\cite{Yuan16SJ,Mertzler14Denoising} that the optimization-based algorithms include two essential steps:
$i$) {\em Gradient descent}: updating the current estimate by extracting more information from the measurement and $ii$) {\em Projection to the signal domain}: confining the result to the desired signal space, which can usually be implemented by denoising. Some algorithms such as ADMM and approximate message passing (AMP) require an correction step. 
This has further been re-purposed and generalized to the {\em Plug-and-Play} (PnP)~\cite{Chan2017PlugandPlayAF} framework. It has derived in~\cite{Liu19_PAMI_DeSCI} that GAP (generalized alternating projection)~\cite{Liao14GAP} can be recognized as a special case of ADMM.  
In the following, we derive the solution based on the ADMM framework to unify PnP, GAP and others. 
By introducing an auxiliary parameter $\vv$, the unconstrained optimization in Eq.~\eqref{Eq:reg} can be converted into
\begin{equation} \label{Eq:ADMM_xv}
({\hat \xv}, {\hat \vv}) = \argmin_{\xv,\vv}  \frac{1}{2}\|\yv - \Phimat \xv\|_2^2 + \lambda R(\vv), {\text{ subject to }} ~\xv = \vv.
\end{equation}  
{ADMM decouples $\{\xv, \vv\}$ in this minimization by introducing another parameter $\uv$, and solves it by the following sequence of sub-problems~\cite{Qiao2020_APLP,Liu19_PAMI_DeSCI}}
\begin{eqnarray}
\xv^{(j+1)} &=&  \argmin_{\xv}  \frac{1}{2}\|\yv - \Phimat \xv\|_2^2 + \frac{\rho}{2} \left\|\xv - \left(\vv^{(j)}-\frac{1}{\rho} \uv^{(j)}\right)\right\|_2^2,  \label{Eq:solvex}\\
\vv^{(j+1)} &=&   \argmin_{\vv} \lambda R(\vv) + \frac{\rho}{2}\left\|\vv - \left(\xv^{(j)}+\frac{1}{\rho} \uv^{(j)}\right)\right\|_2^2, \label{Eq:solvev}\\
\uv^{(j+1)} &=& \uv^{(j)} + \rho (\xv^{(j+1)} - \vv^{(j+1)}), \label{Eq:u_k+1}
\end{eqnarray}
where the superscript $^{(j)}$ denotes the iteration number and $\rho>0$ is another auxiliary parameter (can be set to 1 for the sake of simplicity).
Eq.~\eqref{Eq:solvex} is a quadratic form and has a closed-form solution:
\begin{equation}
{\xv}^{(j+1)}  = \left[\Phimat\ts\Phimat + \rho \Imat \right]\inv\left[\Phimat\ts\yv + \vv^{(j)} -\frac{\uv^{(j)}}{\rho} \right].
\end{equation}
As mentioned before, in SCI, $\Phimat\Phimat\ts$ is a diagonal matrix and we can thus define
\begin{equation}
\Phimat\Phimat\ts = {\rm diag}\left\{\psi_1, \dots, \psi_n\right\}, ~~
\psi_i = \sum_{k=1}^{N_t} D^2_{k,i,i}, ~~~\forall i =1,\dots n {\text{~~with~~}} n = N_x N_y,
\end{equation}
where $D_{k,i,i}$ is the $(i,i)$-th element of $\Dmat_k$ in Eq.~\eqref{Eq:Phimat}.
As derived in~\cite{Yuan16ICIP_GAP}, 
\begin{equation}
{\xv}^{(j+1)}  = (\vv^{(j)} -\frac{\uv^{(j)}}{\rho}) + \Phimat\ts \left[\frac{y_1- \left[\Phimat(\vv^{(j)} -\frac{\uv^{(j)}}{\rho})\right]_1}{\rho + \psi_1},\dots, \frac{y_n- \left[\Phimat(\vv^{(j)} -\frac{\uv^{(j)}}{\rho})\right]_n}{\rho + \psi_n}\right]\ts, \label{Eq:xj+1}
\end{equation}
where $[\av]_i$ is the $i$-{th} element in $\av$ and since $\left\{y_i - \left[\Phimat(\vv^{(j)} -\frac{\uv^{(j)}}{\rho})\right]_i\right\}_{i=1}^n$ can be updated in one shot, ${\xv}^{(j+1)}$ can be solved efficiently.  

Eq.~\eqref{Eq:solvev} is a denoising process of $\vv$ and based on the regularization term $R(\vv)$, various denoising algorithms can be used, for example, the sparsity-based  denoiser~\cite{Yuan14CVPR} and TV based denoising~\cite{Yuan16ICIP_GAP}. 
Advanced denoising algorithms, such as weighted nuclear norm minimization (WNNM)~\cite{Gu14CVPR} has also been used~\cite{Liu19_PAMI_DeSCI}. Most recently, the deep denoiser~\cite{Qiao2020_APLP} has further been used and this leads to the emerging PnP-ADMM framework. Implicitly, we have
\begin{equation}
\vv^{(j+1)}= {\rm Denoise}\left(\xv^{(j+1)}+{1\over \rho }\uv^{(j)} \right). \label{Eq:denoise}
\end{equation}

On the other hand, GAP can be used as a lower computational workload algorithm~\cite{Yuan2020_CVPR_PnP} with the following two steps: 
\begin{eqnarray}
\xv^{(j+1)} &=& \vv^{(j)} + \Phimat\ts(\Phimat\Phimat\ts)\inv (\yv - \Phimat  \vv^{(j)}), \\
\vv^{(j+1)}&=& {\rm Denoise}(\xv^{(j+1)}).
\end{eqnarray}
Clearly, GAP has only two steps while ADMM needs three. Again, due to the diagonal structure of $\Phimat\Phimat\ts$, GAP can be solved very efficiently. It has been proved in~\cite{Liu19_PAMI_DeSCI} that under the noiseless case, GAP and ADMM will lead to the same results, while under the noisy case, ADMM can lead to better results due to its robustness to noise; a geometric explanation has been shown in~\cite{Yuan2020_CVPR_PnP} as well as the convergence results of both ADMM and GAP.

\section{Shallow Learning Based Algorithm \label{Sec:Shallow_Learn}}
The regularization based optimization algorithms were designed to investigate the {\em pre-determined} thus presumed structure (such as sparse or piece-wise constant) of the desired signal. Although various priors can be used, each one has its own pros and cons. On the other hand, learning based reconstruction algorithms learn the structure of the desired signal from the available training data. Examples include the dictionary learning based and Gaussian mixture model (GMM)-based algorithms~\cite{Yang14GMM,Yang14GMMonline,Renna16_GMM}. We term them  ``shallow-learning based algorithms" in this article. Usually, these learning based methods learn an {\em over-complete dictionary} based on HD {\em data patches} and employ CS algorithms to impose sparsity on the coefficients.

In addition to the special structure of the sensing matrix described above, the other benefit of SCI (especially for the video SCI shown in Eq.~\eqref{Eq:Y_sum}) is that the measurement, mask and signal are {\em spatially decoupled} (there is a slight difference in CASSI due to disperser). This means that we can work on 3D patches and these patch-based operations can be performed in parallel.

\subsection{Dictionary Learning}
We assume that a learned (usually over-complete) dictionary $\Psimat$ is obtained from the training data and based on which the signal assumes a sparse representation~\cite{Aharon06TSP}. Recalling Eq.~\eqref{Eq:reg}, now for the $i$-{th} patch (assuming size $P\times P \times N_t$), we aim to solve 
\begin{equation}
{\hat \cv}_i=\argmin_{\cv_i} \frac{1}{2}\|\yv_i - \Phimat_i \Psimat \cv_i \|_2^2+ \lambda\|\cv_i\|_1,
\end{equation}
where $\{\yv_i, \Phimat_i\}$ are the measurement patch and sensing matrix patch for the $i$-th patch; $\lambda$ is again a parameter to balance the two terms and we use $\ell_1$-norm to impose sparsity on the representation coefficients $\cv_i$. After all the patch coefficients $\cv_i$ are estimated, we obtain the desired signal patch  ${\hat \xv}_i = \Psimat \hat\cv_i$ and  then aggregate them back to obtain the 3D cube. A Bayesian dictionary learning method was proposed for spectral SCI in \cite{Yuan15_JSTSP_CASSI_SI}.
It is worth noting that the dictionary $\Psimat$ can be shared across patches but can also be different.

\subsection{Gaussian Mixture Models}
The GMM approach assumes that each patch is drawn from a mixture of Gaussian distributions and by learning these Gaussian components from training data, the desired signal can be solved in closed-form. Specifically,
consider the $i$-th spatial-temporal 3D patch (vectorized) $\xv_i$ being drawn from a GMM: 
\begin{equation}
\xv_i \sim \sum_{k=1}^K \xi_k {\cal N}(\xv_i | \muv_k, \Sigmamat_k),
\end{equation} 
where $\muv_k, \Sigmamat_k$ and $\xi_k$ are the mean, covariance matrix and weight of the $k$-th Gaussian components, respectively ($\xi_k >0$ and $\sum_{k=1}^K \xi_k$ =1). These parameters can be learned based on the training video patches. Note that this is independent from the measurement matrix.  

Concerning the measurement model of the $i$-th patch, we have that $\yv_i = \Phimat_i \xv_i + \ev_i$.
It is assumed that $\ev_i\sim {\cal N}(\ev_i | {\bf 0}, \Qmat)$, where $\Qmat$ is a $P^2 \times P^2$ covariance matrix (recalling that $\yv_i$ and $\ev_i$ are of size $P^2\times 1$). We thus have
\begin{equation}
\yv_i | \xv_i \sim {\cal N}(\yv_i | \Phimat_i \xv_i, \Qmat).
\end{equation}
Applying Bayes' rule results in the posterior~\cite{Chen10_TSP_GMM,Yang14GMMonline}
\begin{equation}
p(\xv_i | \yv_i) = \sum_{k=1}^K \tilde{\xi}_{i,k} {\cal N}(\xv_i | \tilde{\muv}_{i,k}, \tilde{\Sigmamat}_{i,k}), \label{Eq:GMMxiyi}
\end{equation} 
with 
\begin{align}
\tilde{\xi}_{i,k} &= \frac{\xi_k {\cal N}(\yv_i | \Phimat_i \muv_k, \Qmat + \Phimat_i\Sigmamat_k \Phimat_i\ts)}{
\sum_{l=1}^K \xi_l {\cal N}(\yv_i |  \Phimat_i \muv_l, \Qmat + \Phimat_i\Sigmamat_l \Phimat_i\ts)}, \label{Eq:GMM_postxi}\\
\tilde{\Sigmamat}_{i,k} &= (\Phimat_i\ts \Qmat\inv \Phimat_i + \Sigmamat_i\inv)\inv,  \label{Eq:GMM_postsigma}\\
\tilde{\muv}_{i,k} &= \tilde{\Sigmamat}_{i,k} (\Phimat_i\ts \Qmat\inv \yv_i + \Sigmamat_k\inv \mu_k),  \label{Eq:GMM_postmu}
\end{align}
which is also a GMM and {in~\eqref{Eq:GMM_postxi}, ${\cal N}(\yv_i | \Phimat_i \muv_k, \Qmat + \Phimat_i\Sigmamat_k \Phimat_i\ts))$ denotes the  multivariate Gaussian probability density function of $\yv_i$ with mean vector $\Phimat_i \muv_k$ and  covariance matrix $\Qmat + \Phimat_i\Sigmamat_k \Phimat_i\ts$}. 
This analytic inversion expression in~\eqref{Eq:GMMxiyi} on one hand results in efficient reconstruction of every patch. On the other hand, it still needs a moderate computation for each patch: i)  \eqref{Eq:GMM_postsigma} can be pre-computed and \eqref{Eq:GMM_postmu} involves only matrix multiplication, and ii) \eqref{Eq:GMM_postxi} needs matrix inversion for each patch across all $K$ Gaussian components.
In addition, it needs an assumed noise value $\Qmat$, which sometimes needs tuning to get good results.

\subsection{Group-based Sparse Coding}
While the above dictionary learning and GMM methods are performed patch by patch, it has been shown that using {\em patch groups} as basic unit and then imposing sparsity (or low rank) on the patch groups will lead to better results in inverse problems~\cite{Mairal_ICCV09_NLSC,Liu19_PAMI_DeSCI}. The key idea is to exploit the {\em non-local similarity} across images. In the SCI problem considered here, the non-local similarity can also be searched across different frames, thus can lead to even better results with the price of longer running time.

\begin{figure}[htbp!]
	\centering
	{\includegraphics[width=\columnwidth]{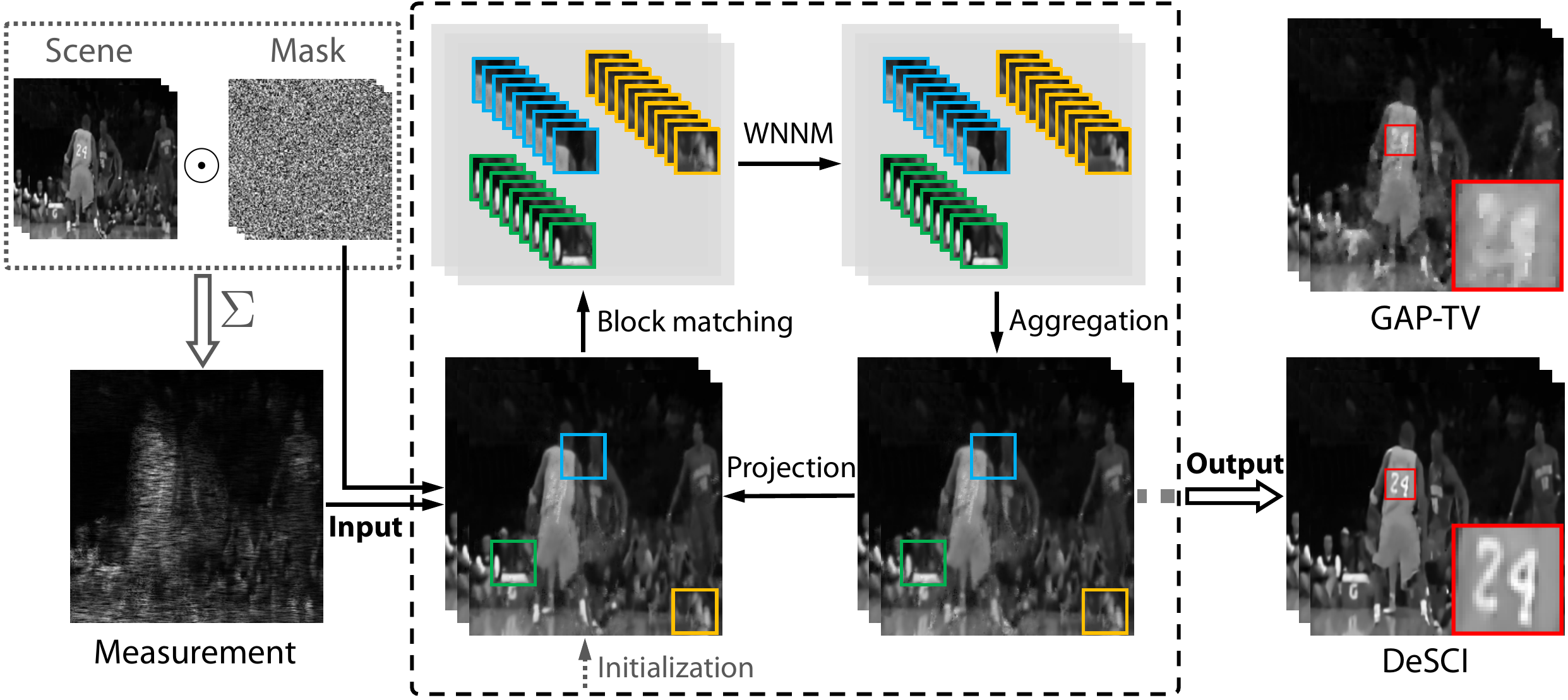}}
	\caption{Flowchart of DeSCI algorithm~\cite{Liu19_PAMI_DeSCI} for SCI reconstruction. Left: the sensing (compressively sampling) process of video SCI. Middle: The proposed rank minimization based reconstruction algorithm, where the projection and weighted nuclear norm minimization (WNNM) for patch groups are iteratively performed.
		Right:  Reconstruction result compared with GAP-TV method~\cite{Yuan16ICIP_GAP}. (Figure from~\cite{Liu19_PAMI_DeSCI}).}
	\label{fig:DescI_flowchart}
\end{figure}

To be concrete, taking the $k$-{th} frame of the desired video in \eqref{Eq:x1toNt}, $\xv_k$ as an example, it is first divided into overlapping patches and each patch is denoted by a vector $\xv_k^j$. Then for each patch $\xv_k^j$, $M$ similar patches are selected from a search window with $C\times C\times T$ (spatio-temporal) pixels to form a set ${\cal S}^j$. Following this, all patches in ${\cal S}^j$ are stacked into a matrix $\Xmat^j = [\xv_k^{j,1}, \xv_k^{j,2},\dots, \xv_k^{j,M}]$. In this matrix, each column represents a patch and since they are searched based on similarity, it will enjoy sparsity under some dictionary and will be low rank. 
Then sparse coding and low-rank algorithms can be performed on these patch groups.
This, along with the ADMM or GAP framework described in Sec.~\ref{Eq:regu_opt} will lead to better results in SCI.  The connection of group sparse coding and low rank models has recently been  made in~\cite{Zha2020_TIP_BenchSC}.
It is worth noting that differently from images, in SCI, each patch can also be 3D and the search window is also 3D. This will enhance sparsity and potentially can further improve the reconstruction results reported in \cite{Liu19_PAMI_DeSCI}, which is depicted in Fig.~\ref{fig:DescI_flowchart}.

\subsection{Summary} 
A short summary of the shallow based learning methods provides the following observations:
\begin{itemize}
	\item Compared with the regularization based algorithms, the patch based learning algorithms can be performed patch-wise and in parallel.
	\item The learning based algorithms are capable of learning complex structures of the signal, yet in patches.
	\item Training data is usually required to learn the structure, though the datasets do not need to be large since the learning is based on patches. Dictionary learning directly from measurements is also feasible but requires iterative algorithms~\cite{Yang14GMMonline,Yuan15_JSTSP_CASSI_SI}. 
	\item In real SCI systems, since the sensing matrix (mask) can not be periodical or a repetitive version of a small mask due to non-uniform illumination patterns and noise, an inherent loop for patches is required to perform these shallow-based patch learning algorithms. 
	\item  The shallow learning algorithms cannot capture the global structure of the signal again because they are based on patches. 
	\item Group-based sparse coding (can be based on dictionary learning or low-rank) algorithms can lead to state-of-the-art results due to the exploration of non-local similarity but with a price paid for running time in patch matching.
	\item In general, since the shallow based learning methods need to divide the images into patches, perform inversion for each patch (group) and then aggregate patches back to images, they cannot lead to real time reconstructions. This represents a significant drawback.
\end{itemize}

In a nutshell, a fast regularization based optimization (\eg, GAP-TV~\cite{Yuan16ICIP_GAP} or TwIST~\cite{Bioucas-Dias2007TwIST}) or shallow learning algorithm (\eg, GMM) cannot lead to outstanding results while a high performance algorithm (\eg, DeSCI shown in Fig.~\ref{fig:DescI_flowchart}) usually needs a long running time due to the extensive block/patch-matching and complicated computation. 


\section{Deep Learning Based Algorithms \label{Sec:Deep}}
\begin{figure}
	\centering
	\includegraphics[width=1\columnwidth]{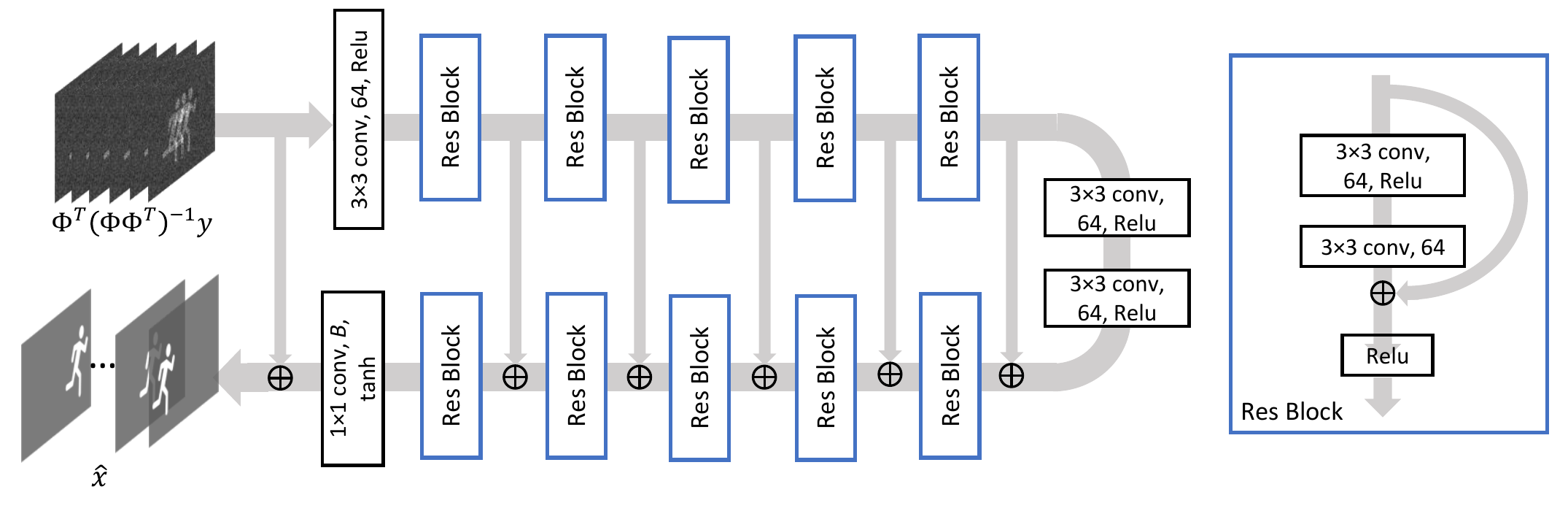}
	\caption{\label{fig:Resnet} Details of the E2E-CNN for video SCI, where $\oplus$ denotes summation. Note the input is now $\Phimat\ts(\Phimat\Phimat\ts)\inv \yv$, which has led to better results than directly input $\yv$ and masks into the network used in~\cite{Miao19ICCV}. (Figure from~\cite{Qiao2020_APLP})}.
\end{figure}

\begin{figure}
	\centering
	\includegraphics[width=1\columnwidth]{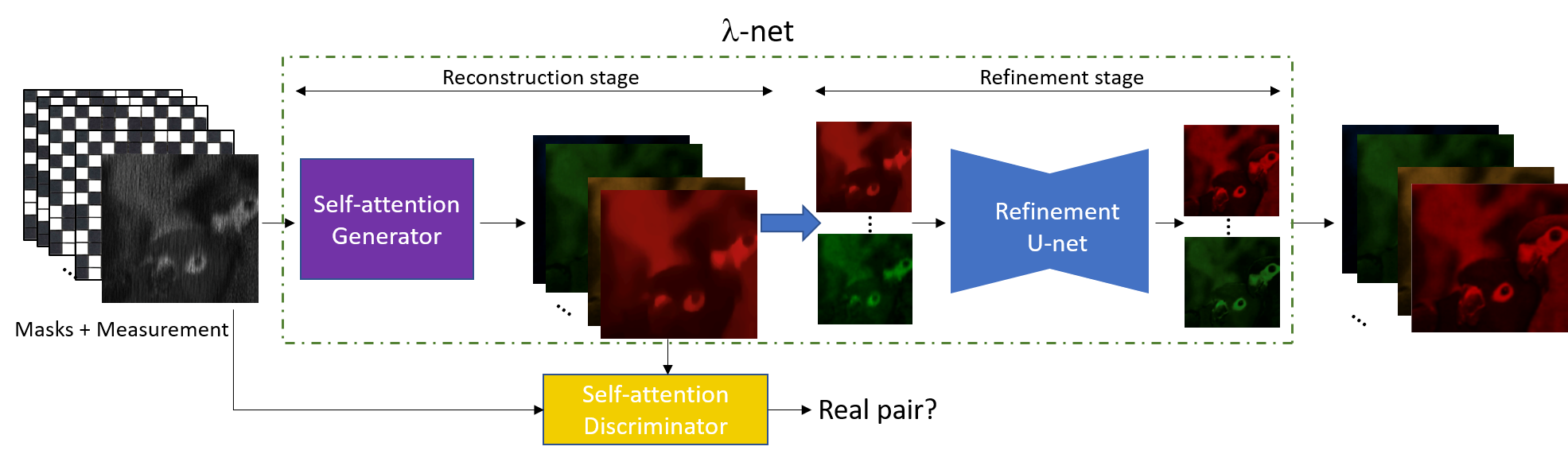}
	\caption{\label{fig:lamnet} Structure of $\lambda$-net (E2E-CNN) for spectral SCI. The masks along with the measurement are fed into the $\lambda$-net to reconstruct the hyperspectral data-cube. Two stages exist in  $\lambda$-net, where the first stage (reconstruction stage) consists of a self-attention GAN plus a hierarchical channel reconstruction strategy to generate the 3D cube from masks plus measurement and the second stage (refinement stage) refines the hyperspectral images in each channel. (Figure revised from~\cite{Miao19ICCV})}.
\end{figure}

Inspired by the powerful learning property of convolutional neural networks, researchers started to employ deep CNNs to learn (an approximation of) the inverse process of the forward model in computational imaging. Specifically, for SCI considered here, the input of a CNN is the measurement $\yv$ (and optionally $\Phimat$) and the output is the desired signal $\xv$. As shown in Fig.~\ref{Fig:DL}(b), this is dubbed end-to-end convolutional neural networks (E2E-CNN) based algorithms. The development of any neural network mostly includes a training phase and testing phases. During training, thousands (or millions) of training pairs $(\yv,\xv)$ are fed into the deep CNN to optimize its parameters. After training, in the testing phase, we only have the measurement $\yv$. By feeding this $\yv$ into the pretrained network, the reconstructed signal $\hat{\xv}$ can be obtained instantaneously since it is just a feed-forward model. 
It has been shown that an encoder-decoder structure with skip connections CNN can usually lead to good results for SCI reconstruction, similar to other computational imaging problems~\cite{Barbastathis19DL}. 
Specifically, a U-net~\cite{Unet_RFB15a} structure by employing the skip connection has been used in~\cite{Qiao2020_APLP} along with residual learning for video SCI as shown in Fig.~\ref{fig:Resnet}.
By integrating the U-net with Generative Adversarial Nets (GAN)~\cite{Goodfellow14_GAN} and the self attention mechanism~\cite{Vaswani17_Attention}, improved results can be achieved. See for example, the $\lambda$-net shown in Fig~\ref{fig:lamnet}, where a two stage network was proposed~\cite{Miao19ICCV}\footnote{Code at: \url{https://github.com/xinxinmiao/lambda-net}}, with the second stage being a refinement network to improve the reconstruction results from the first stage.  

To design an E2E-CNN for SCI reconstruction, one needs to consider the following three aspects:
\begin{itemize}
	\item Network structure: according to our experience, an encoder-decoder structure with skip connections can be used as a backbone. Recurrent neural networks (RNN) can be employed to explore the temporal/spectral correlation within the SCI signals~\cite{Cheng20ECCV_Birnat}\footnote{Code at: \url{https://github.com/BoChenGroup/BIRNAT}.}.
	The attention mechanism can be used to investigate the non-local and high dimensional correlations~\cite{Meng20ECCV_TSAnet}. Integrating attention into a backbone network can lead to better results.    
	It is also worth noting that most of the existing CNNs used for SCI are based on 2D CNNs; 3D (or HD) CNNs can potentially improve the results. 
	The network also needs to scale with the data size. For example, a deeper network is required for a larger scale SCI system.
	\item Loss function: since the main target of SCI is to reconstruct the desired signal, the mean square error (MSE) or $\ell_2$ loss is generally used. The structural similarity, spectral constancy~\cite{Wang19_CVPR_HSSP}, feature loss~\cite{Johnson2016Perceptual} and GAN loss~\cite{Miao19ICCV} can improve the results by providing more details in the reconstructed images. 
	\item Training data: as the forward model in SCI systems can typically be modeled accurately, we can generate simulated measurements using synthetic data. This saves a significant amount of effort required to capture a large number of real data. In particular, it is challenging to obtain the ground truth of high-speed scenes and hyperspectral scenes for SCI systems.
	Therefore, a {\em trained on simulation, testing on real}  data framework is widely used. Lastly, if the data size is too large, \eg, $N_t=50$ in \cite{Qiao2020_APLP}, the training data will be huge and the network needs be very deep; this poses a challenge on GPU memory and training time.
\end{itemize}

In summary, E2E-CNNs enjoy the advantage of fast inference after training. The training time for SCI reconstruction networks depends on the size of the data and usually is on the order of days or weeks while the testing time is usually on the level of tens of milliseconds. 
The disadvantages of E2E-CNNs include the large amount of training data and excessive training time. In addition, E2E-CNN lacks flexibility. Specifically, if the mask or sampling rate (both of them are hardware related) changes, a new network has to be re-trained, which again needs a long time and a large amount of new training data. 
Transfer learning is a good research direction to mitigate this challenge. 
In addition, although we have mentioned that a trained on simulation, testing on real data framework can be used for SCI systems, capturing real data for training will enable CNNs to learn the noise, nonlinearity and uncertainties of the real system. This will make the trained CNNs more robust during reconstruction.

\section{Bridge the Gap Between Deep Learning and Optimization  \label{Sec:bridge}}
Iterative optimization algorithms do not require training data but are usually slow. E2E-CNNs are fast but require long training time. One natural question that arises is whether it is possible to merge the advantages of these two algorithms.
In the literature, two frameworks have been proposed: deep unrolling (unfolding)~\cite{hershey2014deep_unfold} based algorithms (Fig.~\ref{Fig:DL}(c)) and the Plug-and-Play (PnP) based algorithms (Fig.~\ref{Fig:DL}(d)). Differently from E2E-CNN, instead of training a large end-to-end CNN, deep unfolding trains a concatenation of small CNNs (to emulate the iterative operations in traditional optimization), and each of them is called a stage. The optimization based updating rule (linear projection) is employed to connect these stages. Therefore, deep unfolding is an ``unfolding" of a number of iterations of the optimization based algorithm, but it is usually trained in an end-to-end manner. It has some interpretability as each stage corresponds to one iteration. On the other hand, the PnP based algorithms employ pre-trained deep denoising networks as priors and integrate them into the iterative algorithms. They are training free for different applications. Therefore, it is flexible for different applications and systems. It has been proved that given some conditions of the loss function and denoiser, PnP-ADMM converges to a fixed point~\cite{Chan2017PlugandPlayAF}.

\begin{figure}[htbp!]
	\centering
	{\includegraphics[width=1\columnwidth]{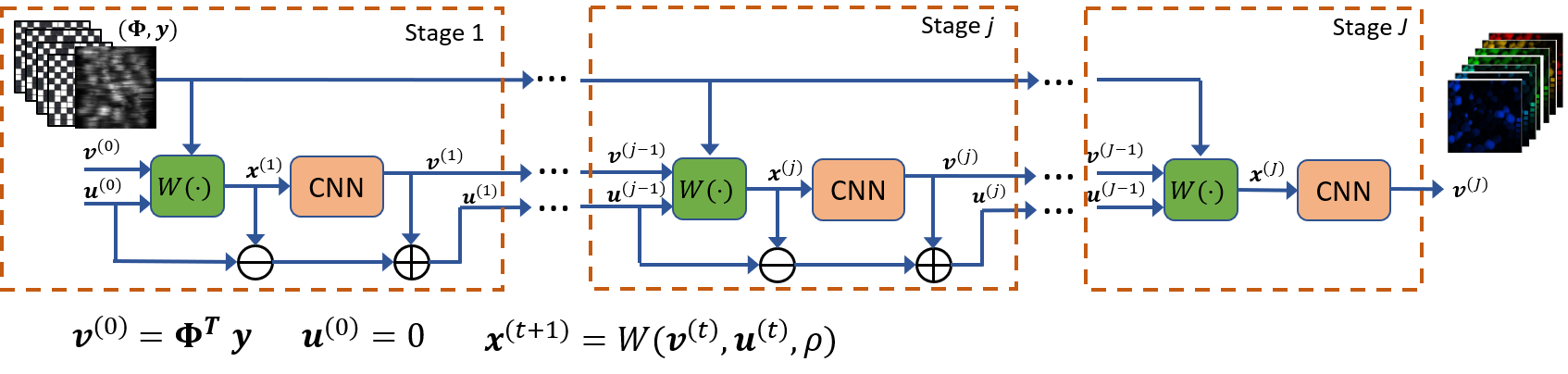}}
	\vspace{-8mm}
	\caption{Structure of ADMM-net (deep unfolding) for SCI reconstruction, where $W(~)$ denotes the computation in \eqref{Eq:xj+1}. Note that these $J$ stages are trained in an end-to-end manner.}
	\label{Fig:admm-net}
\end{figure}

\subsection{Deep Unfolding}
Recalling the ADMM updating equations in \eqref{Eq:solvex}-\eqref{Eq:u_k+1} the denoising step is now replaced by a CNN. This leads to the so called ADMM-net~\cite{Yang_NIP16_ADMM-net} and by using it with tensors applying to SCI results in the deep-tensor-ADMM-net~\cite{Ma19ICCV}\footnote{Code at:\url{https://github.com/Phoenix-V/tensor-admm-net-sci}.}.
Without explicitly showing the CNN being used, we depict the ADMM-net for SCI in Fig.~\ref{Fig:admm-net}. As mentioned before, each stage corresponds to one iteration in the optimization based ADMM algorithm; however, here we only need a dozen stages and within each stage, there is one CNN (playing a similar role to denoising) and these $J$ stage CNNs are trained in an end-to-end manner.
Even though the structure of the CNNs can be the same, each of them is  playing a different role and obviously, the learned parameters (weights in the CNNs) are different. 
Intuitively, the first few stages need to denoise an image with a high noise level while the last few stages will refine the signal to retrieve details. This is the interpretability of the deep unfolding approach mentioned in Sec.~\ref{Sec:Motivation}.
This shares the similar spirit of the following Plug-and-Play framework.

Although the deep unfolding network is trained in an end-to-end manner, which is similar to  the training of an E2E-CNN, the difference is significant, since the small CNNs are independent from the sensing matrix $\Phimat$, and thus they can be trained/performed with a smaller dimension (\eg, blocks or patches) than the size of the desired signal. In this case, both their training and testing might be faster than their counterpart of E2E-CNN, which inputs the measurement $\yv$ plus the sensing matrix $\Phimat$ and outputs the signal $\hat{\xv}$ directly.

One potential problem with deep unfolding is the number of total stages $J$.
The theoretical analysis of PnP in~\cite{Yuan2020_CVPR_PnP} can also be utilized here and the reconstruction error is related to the iteration number.
Specifically, if $J$ is large, it is challenging to train the network in an E2E manner; on the other hand, if $J$ is small, the results tend to be degraded. 
In addition, after training, each CNN is fixed and the whole network might not be robust to noise. In the literature, deep unfolding networks have  achieved good results with simulated data, since the noise model is the same both during training and testing. However, with real data, the performance degrades and sometimes they perform worse than E2E-CNNs. Furthermore, they are not as flexible as the following PnP based algorithms.
Most recently, the GAP-net proposed in~\cite{Meng_GAPnet_arxiv2020} has achieved good results in both video and spectral SCI. A deep unfolding based on Gaussian scale mixture model has been developed in~\cite{Huang2021_CVPR_GSMSCI} for spectral SCI reconstruction\footnote{Code at:\url{https://github.com/TaoHuang95/DGSMP}.}.

\subsection{Plug-and-Play \label{Sec:PnP}}
As mentioned before, the PnP framework~\cite{Chan2017PlugandPlayAF} replaces the denoising step in the optimization based algorithm with a {\em deep denoising prior} or a {\em deep denoiser}. One key property of the deep denoiser is that it needs to be robust to noise, \ie, denoising the image with a wide range of noise level. Fortunately, recent advances in deep denoising networks have provided us with plenty of well pre-trained neural networks, for example the FFDNet~\cite{Zhang18TIP_FFDNet}. Applying the FFDNet with ADMM or GAP into SCI has led to fast, flexible and efficient algorithms~\cite{Yuan2020_CVPR_PnP}. 
The PnP algorithm is efficient due to the fact that it consists of the same 3 steps in ADMM (2 steps in GAP) as in the optimization method; when the denoising step is replaced by the deep denoiser, it is very fast.
Usually, less than 100 iterations are sufficient to provide good results in video SCI systems.
Moreover, the convergence results~\cite{Ryu2019PlugandPlayMP} can be obtained by appropriate assumptions, \eg, the bounded gradient of the loss function and the bounded denoiser used in~\cite{Chan2017PlugandPlayAF}.
This convergence is the {fixed point convergence} based on ADMM and it is easy to verify that in the SCI forward model, the loss function $f(\xv) = \frac{1}{2}\|\yv- \Phimat \xv\|_2^2$ has bounded gradient and thus PnP-ADMM for SCI also enjoys the fixed point convergence. 
Most recently, the PnP-GAP has been developed for SCI with proven  convergence~\cite{Yuan2020_CVPR_PnP}.
It is not necessary to reproduce the theoretical results here and the key point is that the reconstruction error depends on the performance of the bounded denoiser and with the iteration number increasing, the result is getting better.
This works in concert with the second and third challenges posed in Sec.~\ref{Sec:Motivation}.
  
A significant advantage of PnP is that it uses a pre-trained network. This also leads to the second advantage of being flexible, \eg, it is robust to encoding masks, sampling rates, \etc.
By integrating PnP-ADMM or PnP-GAP into flexible deep denoisers, the results will be comparable with  state-of-the-art approaches. As mentioned above, since we only need less than 100 iterations for PnP to give good results, it is therefore very efficient. Regarding flexibility, it has been shown to achieve excellent results on different sampling rates, masks and real video SCI cameras~\cite{Yuan2020_CVPR_PnP} by using the same deep denoiser. 
Therefore, PnP provides a good trade-off among speed, accuracy and flexibility.
We recommend this as the new baseline for SCI reconstruction. 
A joint reconstruction and demosaicing framework has recently been proposed in~\cite{PnP_SCI_arxiv2021} for video SCI\footnote{Code at: \url{https://github.com/liuyang12/PnP-SCI_python}}.

One potential problem of PnP for SCI is that in some imaging systems, \eg, hyperspectral images, a flexible and efficient denoiser does not exist. In this case, a new network needs to be trained before we can plug it into the PnP framework. Yet, training an efficient and flexible hyperspectral deep denoising network is challenging and the ultimate performance of the PnP relies heavily on the performance of the deep denoiser and the denoiser in~\cite{Zheng20_PRJ_PnP-CASSI} has shown high performance for spectral SCI\footnote{Code at: \url{https://github.com/zsm1211/PnP-CASSI.}}.

\subsection{Scale to Large Dataset}
It has been demonstrated in~\cite{Yuan2020_CVPR_PnP} that PnP algorithms can easily scale up to large dataset such as 4K videos.
However, due to the iterative property, PnP will be slow for large-scale data, \ie, far from real time.
As shown in Table~\ref{Tab:Algo_comp}, for the sake of high speed, we need E2E-CNN or deep unfolding.
However, this brings the challenges of big model size and large training dataset. To address these issues, two solutions are proposed in the literature:
\begin{itemize}
    \item The first one is to train a small network but to adapt it to different modulation masks. One recent work has been done in~\cite{Wang2021_CVPR_MetaSCI} demonstrating the promise of this direction using meta learning\footnote{Code at: \url{https://github.com/xyvirtualgroup/MetaSCI-CVPR2021.}}.
    \item The other solution is to develop memory efficient networks using techniques such as reversible network~\cite{Cheng2021_CVPR_ReverSCI}\footnote{Code at: \url{https://github.com/BoChenGroup/RevSCI-net.}}.
\end{itemize}

\subsection{Untrained Network: Deep Image Prior}
A most recent method to combine optimization and deep learning is the {\em deep image prior}~\cite{Lempitsky2018CVPR_DIP}, which employs an {\em untrained} neural network to learn the prior directly from the degraded image. This is a good research direction in SCI reconstruction.
However, since it still needs (usually on the order of thousands of) iterations to optimize the parameters in the network, the computational workload is high and thus the algorithm will be slow. 
Some preliminary results of video SCI are shown in~\cite{Qiao2021_MicroCACTI}.

\subsection{Summary}
Both deep unfolding and PnP have been used into different SCI systems for reconstruction.
Table~\ref{Tab:SCI_algo} lists the various algorithms used in various SCI systems. 
At the time of writing this manuscript, the authors are not aware of any work using the deep image prior for SCI reconstruction.

\begin{itemize}
	\item Deep unfolding is composed of $J$ small CNNs and each CNN plays the role of a denoiser. From this perspective, it can be very fast. However, as mentioned before, the performance of the network depends on the number of stages being used. Also, each CNN cell needs to be deep to achieve good results in which case it is not fast anymore. Additionally, the network is not robust to real data noise. In principle, deep unfolding networks are flexible, meaning that they can be used for different sensing matrices. However, in realistic cases, since they are trained in an end-to-end manner, it is difficult to achieve competitive results on an SCI system by using the network trained on a different one.
	\item PnP employs a pre-trained deep denoiser (neural network) and it provides a good trade-off among speed, accuracy and flexibility. A good deep image denoiser can be used into different SCI systems to achieve good results. 
	However, devising a good deep denoiser might be challenging for some SCI signals. A spectral denoiser has been trained in~\cite{Zheng20_PRJ_PnP-CASSI} and achieved good results for spectral SCI\footnote{Code at: \url{https://github.com/zsm1211/PnP-CASSI}}.
	\item  The use of a deep image prior might be a good research direction for SCI when abundant training data are not available but speed will be an issue. 
\end{itemize}

In general, the algorithm aims to solve the {\em trilemma of speed, accuracy and flexibility} for SCI reconstruction. Table~\ref{Tab:Algo_comp} lists these properties of different algorithms. 

\begin{table}[!htbp]
	\caption{Property (speed, accuracy and flexibility) of different algorithms for SCI reconstruction. H: High, M: Middle, L: Low.}
	\centering
	{
		\begin{tabular}{|c || c|c|c|}
			\hline 
			Algorithm& Speed & Accuracy & Flexibility\\
			\hline \hline
			Conventional Regularization Optimization         & L & H/M/L & H\\
			\hline
			Shallow-learning  & M & M/L & H/M \\
			\hline 	\hline
		    E2E-CNN      & H& H/M& L  \\
			\hline
			Deep Unfolding      & H & H & M \\
			\hline
			Plug-and-Play   & H/M& H & H \\
			\hline
	\end{tabular}}
	\label{Tab:Algo_comp}
\end{table}

\section{Results of Experimental Data \label{Sec:Real}}

\begin{figure}[htbp!]
	\centering
	{\includegraphics[width=1.0\textwidth]{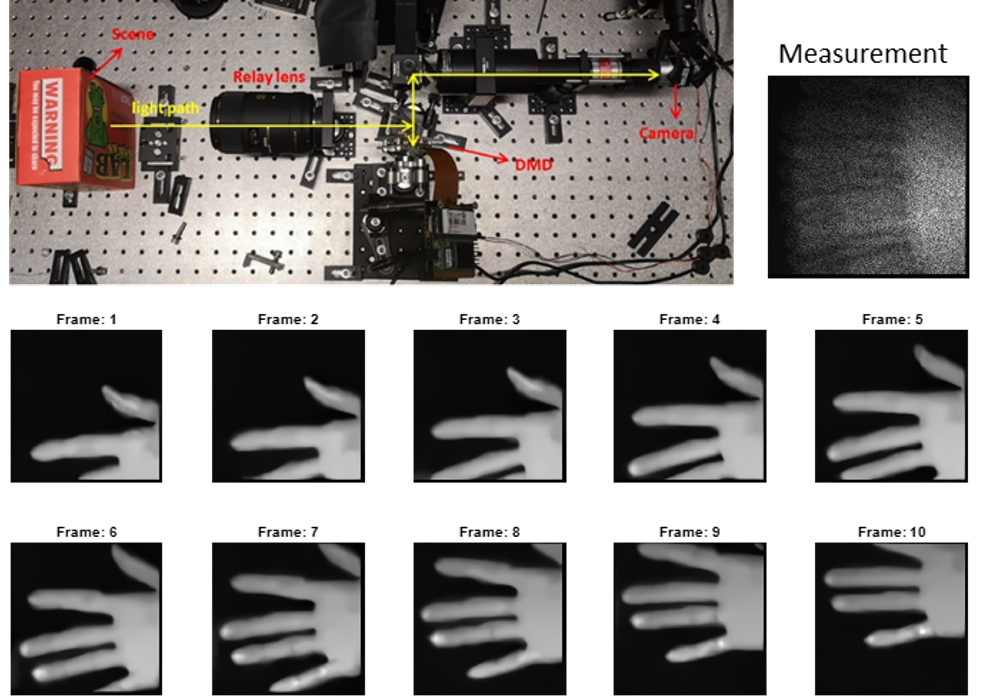}}
	\caption{Video SCI system (top-left) built at Bell Labs and one measurement (top-right) as well as 10 corresponding reconstructed frames (bottom, reconstructed by PnP-FFDnet~\cite{Yuan2020_CVPR_PnP}).}
	\label{Fig:cacti_sys}
\end{figure}

In this section, we use two SCI systems built at Bell Labs, one for video SCI shown in Fig.~\ref{Fig:cacti_sys} and the other one for spectral SCI shown in Fig.~\ref{Fig:cassi_sys}, to compare different algorithms based on real data captured by these systems. 
For a quantitative comparison, simulation results based on the benchmark synthetic datasets are also  briefly presented.    

\subsection{Video SCI System}

\begin{figure}[htbp!]
	\centering
	{\includegraphics[width=1\columnwidth]{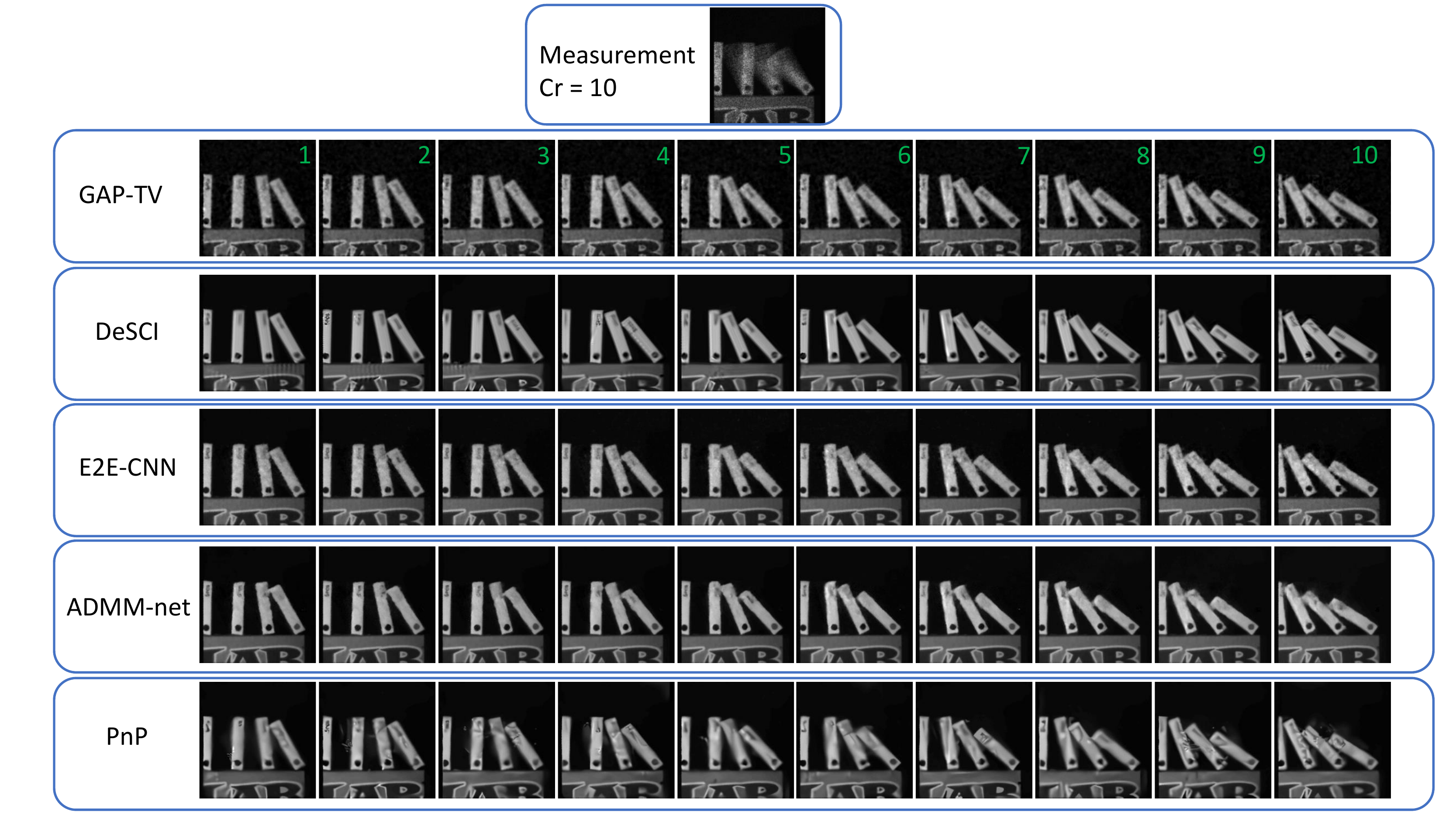}}
	\caption{Real data results of video SCI at Cr = 10, \ie, 10 frames (marked by green numbers in the plot) reconstructed from a single measurement (top) captured by our video SCI camera in Fig.~\ref{Fig:cacti_sys}. These 10 frames happen within 20ms in real life and this leads to a high-speed video camera at 500 frame-per-second (fps) by using a camera working at 50fps.}
	\label{Fig:cacti_cr10}
\end{figure}

The optical setup of our video SCI system is the one reported in~\cite{Qiao2020_APLP}, depicted in Fig.~\ref{Fig:cacti_sys}. A commercial photographic lens (L1) images the object onto an intermediate plane, where a DMD (Vialux, DLP7000) is employed to apply (random binary) spatial modulation to the high-speed image sequences. A 4f system consisting of a tube lens L2 (f=$150mm$) and an objective lens ($4\times$, f=$45mm$, NA=$0.1$) relays the modulated image to a camera (Basler, acA1300-200um). 
The camera operates at a fixed frame rate of $50$ frames per second (fps), whereas the DMD operates at various frame rates between $500$ and $2500$ fps, resulting in compressive ratios (Cr) ranging from $10$ to $50$ (for clarification, Cr=$10$ means $10\times$ compression, \ie, $N_t = 10$.  The camera and the DMD are synchronized by a data acquisition board (NI, USB6341), so that when the DMD changes Cr ($N_t$) times, the camera captures one 2D measurement. The modulation patterns used in the reconstruction are pre-calibrated by illuminating the DMD with a uniform light and capturing the images of the DMD patterns with the camera.

\begin{figure}[htbp!]
	\centering
	{\includegraphics[width=1\columnwidth]{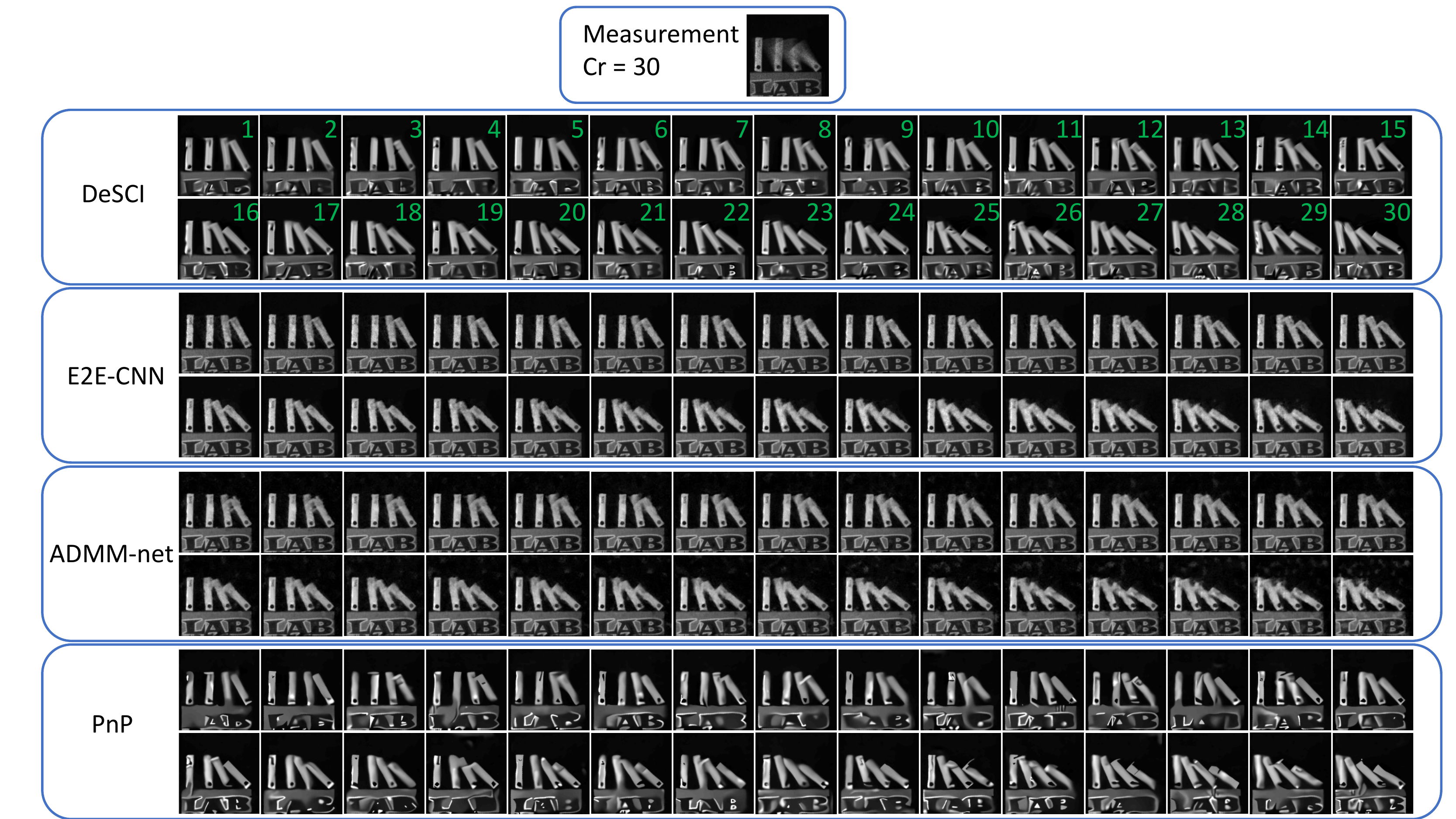}}
	\caption{Real data results of video SCI at Cr = 30, \ie, 30 frames (marked by green numbers in the plot) reconstructed from a single measurement (top) captured by our video SCI camera in Fig.~\ref{Fig:cacti_sys}. Note that these 30 frames happen within 20ms in real life and this leads to a high-speed video camera at 1500 fps (by using a camera working at 50fps).}
	\label{Fig:cacti_cr30}
\end{figure}

\subsection{Experimental Results of Video SCI}
In this section, we selected one high-speed scene \texttt{falling dominoes} captured by our video SCI system and show the results at Cr=10 and Cr=30 using different algorithms, namely, DeSCI~\cite{Liu19_PAMI_DeSCI}, E2E-CNN~\cite{Qiao2020_APLP}, deep unfolding (ADMM-net) and PnP-FFDnet~\cite{Yuan2020_CVPR_PnP}.
Note that we re-purposed the ADMM-net for video SCI, where in each stage, a U-net is used; it is trained on the same datasets as the E2E-CNN in~\cite{Qiao2020_APLP}\footnote{More data and results are available at \url{https://github.com/mq0829/DL-CACTI}.}.
The measurement size is $512\times 512$ and the reconstructed frames at Cr=10 are shown in Fig.~\ref{Fig:cacti_cr10}, where we can see that DeSCI\footnote{Code at:\url{https://github.com/liuyang12/DeSCI}}, E2E-CNN, ADMM-net\footnote{Code at:\url{https://github.com/mengziyi64/ADMM-net}} (deep unfolding) and PnP\footnote{Code at: \url{https://github.com/liuyang12/PnP-SCI}} can all provide good results and they are better than GAP-TV, which is a baseline using total variation. 
Therefore, we further show a challenging case at Cr=30, \ie, 30 frames are reconstructed from one measurement in Fig.~\ref{Fig:cacti_cr30}. We can see now that PnP starts to having unpleasant artifacts as well as DeSCI (please refer to `LABS' at the bottom part). However, DeSCI and PnP can provide clear motion details while E2E-CNN and ADMM-net lead to blurry results.

\begin{table*}[!htbp]
	\caption{Average PSNR result in dB (left entry in each cell) and SSIM (right entry in each cell) and running time for 6 benchmark simulation datasets for video SCI. A group of \{ground truth, measurement\} is called a ``pair" in the training data.}
	\centering
	{
		\begin{tabular}{c||cc|ccc}
			\hline
			Algorithm& GAP-TV~\cite{Yuan16ICIP_GAP} & DeSCI~\cite{Liu19_PAMI_DeSCI} & E2E-CNN~\cite{Qiao2020_APLP} & PnP~\cite{Yuan2020_CVPR_PnP} & Deep Unfolding (ADMM-net)  \\
			\hline
			Average results & 26.73, 0.858& 32.65, 0.935 & 29.59, 0.900 & 29.70, 0.892 & 32.53, 0.943    \\
			Running time (sec) &4.2 (CPU)& 6180 (CPU) & 0.023 (GPU) & 3.0 (GPU) & 0.058 (GPU)\\
			Training time &-& - & 33 hours & - & 120 hours\\
		   Training data size &-& - & 26,000 pairs& - & 26,000 pairs\\
			\hline
	\end{tabular}}
	\label{Tab:results_video}
\end{table*}

For a quantitative comparison, simulation results based on the benchmark synthetic data are presented here.    
The datasets include six high-speed scenes, where $N_t=8$ video frames with size $256\times 256$ are compressed into a single measurement. We use the same measurement as in~\cite{Yuan2020_CVPR_PnP}.
Table~\ref{Tab:results_video} summarizes the PSNR and SSIM~\cite{Wang04imagequality} results of these 6 benchmark datasets.
It can be seen that DeSCI still leads to the highest PSNR and SSIM, and ADMM-net is the runner-up.
From the speed perspective, E2E-CNN and ADMM-net are the fastest. 

\begin{figure}[htbp!]
	\centering
	{\includegraphics[width=1\columnwidth]{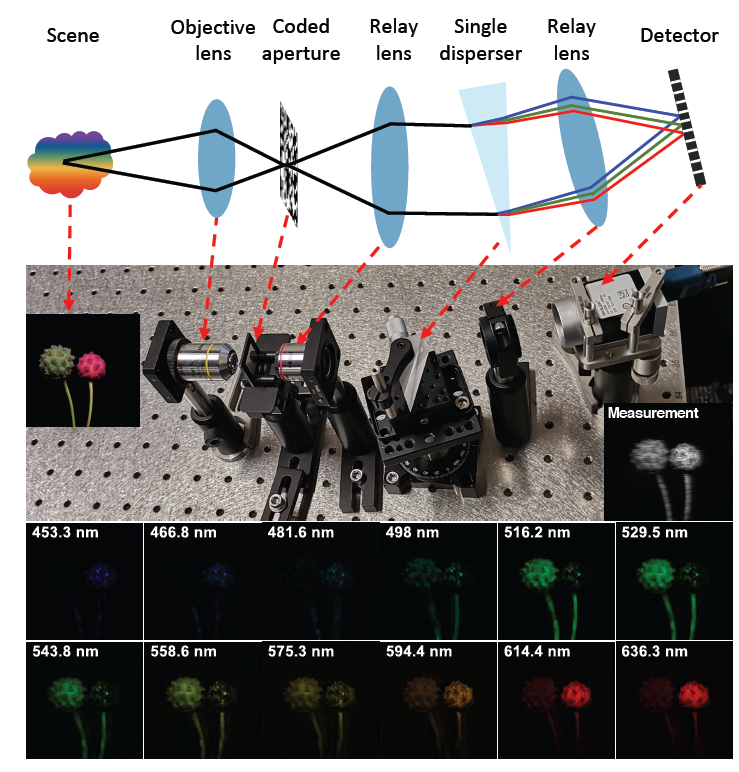}}
	\caption{Spectral SCI system (top-left) built at Bell Labs and one measurement (bottom-right of the upper part) and reconstructed hyperspectral images shown 12 different wavelengths (lower part, reconstructed by E2E-CNN).}
	\label{Fig:cassi_sys}
\end{figure}

\subsection{Spectral SCI System}
The hardware system of our spectral SCI is the re-built single disperser CASSI reported in~\cite{Meng20ECCV_TSAnet}. As shown in Fig.~\ref{Fig:cassi_sys}, it consists of a 18mm objective lens (Olympus RMS10X), a coded aperture (mask), two relay lenses (Olympus RMS4X 45mm and Thorlabs AC254-050-A-ML 50mm), a dispersive prism (Edmund 43649) and a detector (Basler acA2000-340km). The detector has $2048\times 1088$ pixels with the pixel pitch of 5.5$\mu$m and supports 12-bit depth.
The system is calibrated by a laser to capture the mask used in reconstruction. Two edgepass filters are used to limit the spectral range of the system to 450-650 nm. In this range, the prism with $30^\circ$ apex produces 54-pixel dispersion. By calibrating the prism, we determine 28 spectral channels at wavelengths: \{453.3, 457.6, 462.1, 466.8, 471.6, 476.5, 481.6, 486.9, 492.4, 498.0, 503.9, 509.9, 516.2, 522.7, 529.5, 536.5, 543.8, 551.4, 558.6, 567.5, 575.3, 584.3, 594.4, 604.2, 614.4, 625.1, 636.3, 648.1\}nm.


\begin{figure}[htbp!]
	\centering
	{\includegraphics[width=1\columnwidth]{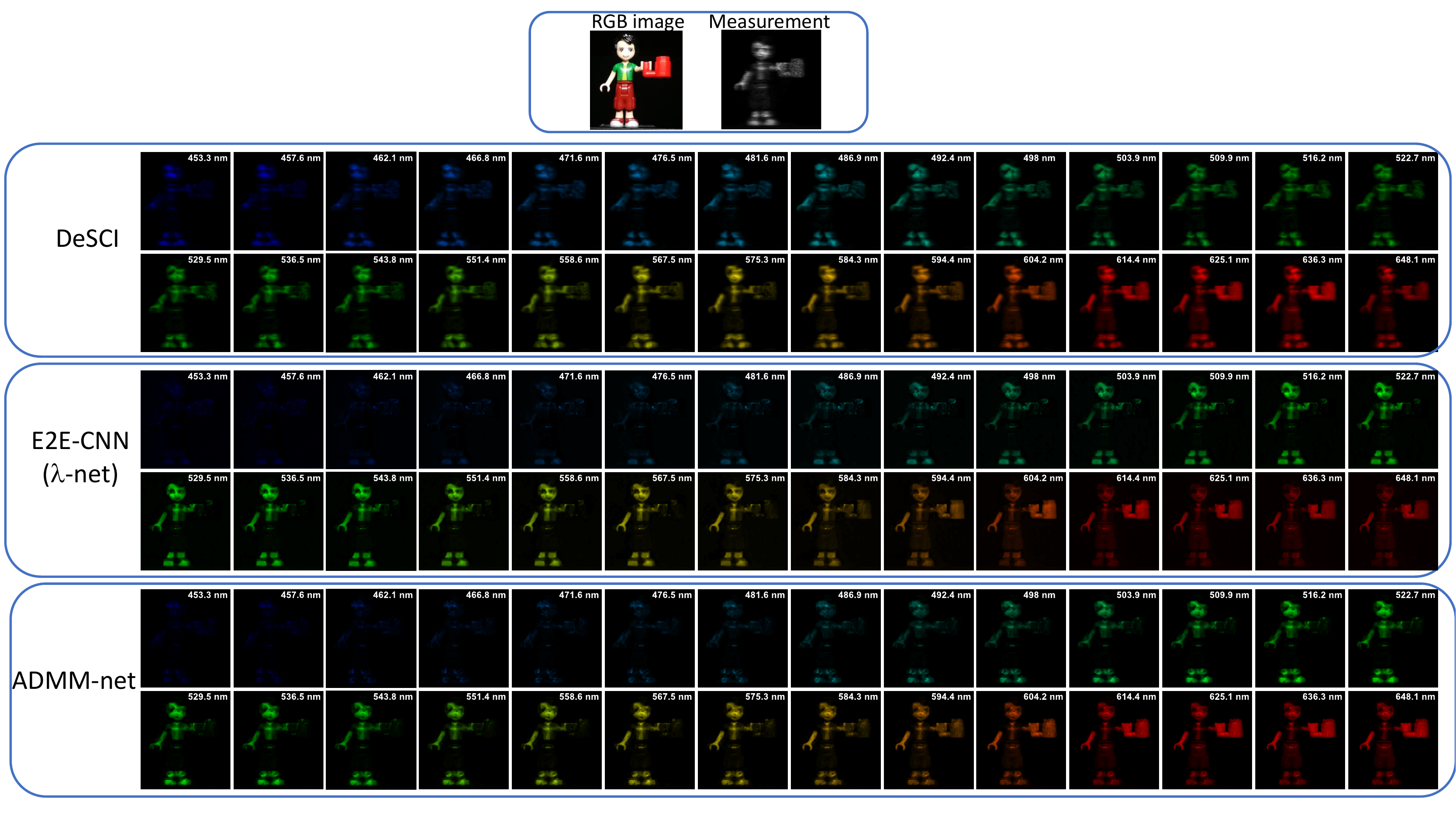}}
	\caption{Real data results of spectral SCI. 28 frames (at different wavelengths, marked on the top-right in each frame) are reconstructed by different algorithms from a single measurement (top-right) captured by our CASSI system in Fig.~\ref{Fig:cassi_sys}. The RGB image is shown on the top-left for reference. Note that the red part should be strong at channels with wavelength larger than 600nm and green part should be strong between 520nm to 560nm wavelengths.}
	\label{Fig:cassi_real}
\end{figure}

\subsection{Experimental Results  of Spectral SCI}
Fig.~\ref{Fig:cassi_real} shows one exemplar result of a spectral scene captured by our spectral SCI camera using different algorithms, namely DeSCI, E2E-CNN ($\lambda$-net) and our re-purposed ADMM-net.  
We can see that both $\lambda$-net and ADMM-net lead to better results than optimization based algorithms. We did not show results of other optimization based algorithms here since they are inferior to DeSCI.    
It can be observed that DeSCI leads to blurry results while the contrasts of $\lambda$-net and ADMM-net are higher\footnote{More data and results are available at \url{https://github.com/mengziyi64/TSA-Net}.}.

\begin{table*}[!htbp]
	\caption{Average PSNR result in dB (left entry in each cell) and SSIM (right entry in each cell) and running time for 10 selected simulation scenes for spectral SCI. A group of \{ground truth, measurement\} is called a ``pair" in the training data.}
	\centering
	{
		\begin{tabular}{c|| cc|cc}
			\hline
			Algorithm& GAP-TV~\cite{Yuan16ICIP_GAP}  & DeSCI~\cite{Liu19_PAMI_DeSCI} & E2E-CNN ($\lambda$-net~\cite{Miao19ICCV}) & Deep Unfolding (ADMM-net)  \\
			\hline
			Average results& 23.73, 0.683 & 25.86, 0.785 & 29.25, 0.886 & 32.13 0.924    \\
			Running time (sec) & 35 (CPU) & 10,800 (CPU) &  0.065 (GPU)  & 0.081 (GPU)\\
			Training time & - & - & 48 hours  &  60 hours \\
			Training data size &-& - &  4,000 pairs & 4,000 pairs\\
			\hline
	\end{tabular}}
	\label{Tab:results_spectral}
\end{table*}

For quantitative analysis, we conducted a simulation with 10 scenes from the KAIST datasets~\cite{choi2017high}. 
The measurements are generated following the optical path of our CASSI system shown in Fig.~\ref{Fig:cassi_sys}. 
The simulation is noise free and the average results are summarized in Table~\ref{Tab:results_spectral}.
Note that the $\lambda$-net is a two stage network. 
We can see that the deep learning algorithms perform significantly better than optimization based algorithms with a large margin (over 3dB in PSNR). ADMM-net performs the best in simulation.


\subsection{Summary}
From the results of both real data and simulation, we have the following observations.
\begin{itemize}
	\item For video SCI, DeSCI still leads to the best results, \ie, highest PSNR and SSIM in simulation data and more details in real data. However, it is very slow, \ie, requiring hours to reconstruct the video frames from a single measurement. Among deep learning algorithms, the deep unfolding achieves higher PSNR than E2E-CNN and PnP; this gain also manifests itself in the real data results. E2E-CNN may need more complicated networks such as RNN and self attention mechanism to improve the results (Sec.~\ref{Sec:Deep}) and PnP may need a better video denoiser (rather than FFDnet, which is based on single images) to get competitive results to deep unfolding. 
	However, from the flexibility perspective, PnP is the favorite choice since it does not need re-training. Therefore, we recommend PnP as a baseline for SCI problems if a good denoiser exists.
	\item For spectral SCI, deep learning leads to significant gains over conventional methods in addition to the much shorter running time (after days or weeks of training).   
	Since we have not found a flexible denoising algorithm for spectral images (though we have  tried to train one but the results were not good), we did not show the results of PnP.  
	Although deep unfolding can lead to a higher PSNR on simulation, visually, the real data results from E2E-CNN have the highest quality, \ie, providing more details and fewer artifacts than ADMM-net in Fig.~\ref{Fig:cassi_real} (please refer to the face part). 
\end{itemize}

\section{Applications of Snapshot Compressive Imaging \label{Sec:Appli}}
SCI systems, from their inception, are aiming to capture more information using low cost imaging systems.
Therefore, it is fair to claim that SCI is an application driven research topic. 
Different SCI systems can be used in different environments in our daily life. In the following, we give some examples. 
\begin{itemize}
	\item[i)] Due to the high capturing speed (but low-cost) and the near real-time reconstruction (by the E2E deep learning methods) of video and depth SCI, we expect the end-to-end system to find applications in traffic surveillance, sports photography and autonomous vehicles. Furthermore, these SCI systems can be deployed with edge cloud/computing~\cite{Liu_2019PIEEE_Edge} to lower the bandwidth and memory requirements. 
	Specifically, a light weight CNN can be deployed at the edge (end-users) for a fast detection, while a complicated CNN can be deployed on the central cloud for reconstruction and refinement detection. 
	\item[ii)] Spectral SCI is essentially a low-cost spectrometer, which will find wide applications in public health\footnote{An SCI  endomicroscopy has been built in~\cite{Meng2020_OL_SHEM} with code and data available at: \url{https://github.com/mengziyi64/SMEM}}, chemical engineering and gemological engineering, among others.
	\item[iii)] Microscopy SCI, including optical coherence tomography SCI can be used in biomedical imaging and industrial deficient detection for high-speed and large-area imaging.
	\item[iv)] Volumetric and holographic SCI can be applied to virtual reality, augmented reality and mixed reality.
	\item[v)] X-ray SCI systems including transmitting and scattering SCI can be applied to chemistry material analysis, and airport security inspection, among others.
\end{itemize}
In short, SCI systems can potentially be used into any environment that a conventional (non-SCI) imaging system has been dedicated to.

\section{Outlook of Snapshot Compressive Imaging \label{Sec:Outlook}}
It can be seen from Tables~\ref{Tab:SCI_system} and~\ref{Tab:SCI_algo} that both hardware and algorithms have not been fully developed for various SCI systems. Therefore, new hardware systems and algorithms will be developed as well as new theoretical analysis. 
We anticipate that with the advances in deep learning, various end-to-end SCI systems will be used in our daily life for diverse applications discussed in Sec.~\ref{Sec:Appli}. For instance, the low-cost spectrometer based on spectral SCI can be integrated into our cellphones. This, along with specific task driven software, will help us exploit the environment (\eg, food ingredient analysis and material detection) and thus upgrade our life quality.
In the following, we discuss several challenges that need to be solved for daily applications of SCI and also some future research directions.

\subsection{Dynamic Range of Sensors \label{Sec:dynamicRange}}  
The dynamic range of the sensor is a limitation of the SCI system since in our current setup, each pixel of the detector captures a modulated summation of pixel values across the spectrum and/or time and/or other dimensions. After reconstruction, each pixel may only have a dynamic range of 4 bits or even smaller given an 8 bit camera. Therefore, a high dynamic range camera is desired for SCI systems and we have seen the improved results by replacing the 8 bit camera with a 12 bit one.
This problem might be alleviated by using advanced deep learning algorithms, as they are capable of learning complex structures from the training data.

From the hardware side, one limitation is that we can not implement a mask with negative values in optical imaging to eliminate the DC value. 
One solution is to build a dual-path SCI system with two cameras and two conjugate masks as mentioned in the limitations of the theory in Sec.~\ref{Sec:Theory}.
In this way, we can capture two conjugate measurements simultaneously and using the difference of these two measurements, we obtain the equivalent of a single measurement corresponding to the mask with $\{1,-1\}$. However, for each measurement, we still have the dynamic range problem and thus this does not completely solve the problem . Another advantage of this solution is that this is closer to the theoretical analysis, which usually assumes a zero-mean modulation. On the other hand, this leads to the drawback of higher cost since more complicated hardware will be involved.

\subsection{Power Consumption of Active Modulation}
Modulation is the key in SCI as depicted in Fig.~\ref{Fig:sci_encoder}. 
The elegant design of CASSI (spectral SCI) used passive modulation and thus it is low-cost and no additional power is required; the path to integrate it into mobile devices is clear (by making it small). 
However, for other active modulations in SCI (for instance the video SCI), implementing it into our daily life, especially on mobile devices and autonomous vehicles may raise the concern of power consumption. Therefore, the gain of bandwidth, memory and information compared to the power consumption establishes a trade-off. 
One research direction is to use new materials such as metamaterials~\cite{Cui2014_LSA_meta} and metasurface to implement the modulation, but the image-space coding described in Sec.~\ref{Sec:hard_summary} may be a concern since a pixel-to-pixel correspondence is challenging.

\subsection{Mask Optimization}
Mask optimization is a research direction~\cite{Iliadis_DSP2020} for theoretical analysis, but so far we have not seen a significant improvement compared with random binary masks in {\em real} SCI systems, although it does show a gain in simulation. 
In general, a random binary mask can lead to decent results for uncertain scenes and this usually  serves as a baseline in various SCI systems.   
Recalling the forward model of SCI in Eq.~\eqref{Eq:Phimat}, limited elements in the sensing matrix can be optimized compared to the single-pixel camera system. In other words, there are more strict constraints in SCI sensing matrix. 
Some of the challenges to utilize optimized masks are listed as follows:
\begin{itemize}
	\item The dynamic range of the mask being {\em manufactured} is limited. This is different from the aforementioned dynamic range of the camera. Usually, during mask optimization, the mask is assumed to be a continuous value (single or double format). Binary masks can also be optimized but usually with limited gain. When the optimized mask is {\em  quantized} and manufactured, the gain is shrunk.   
	\item Even though an optimized mask can be designed, there is still {\em fabrication error} and this error might cancel out the gain. The mask is calibrated based on an ideal illumination and then used in the reconstruction. This calibrated mask is different from the one being designed.
	\item Non-ideal illumination exists in real SCI systems. In real cases, the illumination of the scene might not be uniform and this will lead to another error in the final reconstruction.
\end{itemize}
To summarize, there is a gap between theoretical analysis (simulation) and real hardware for mask optimization. One possible solution is to use deep learning to perform join reconstruction and mask optimization as in~\cite{Iliadis_DSP2020} but with mild conditions. Instead of per image optimization, the mask can be optimized for each class of the desired signal, for example, one for faces used in person identification and the other one for cars used in traffic surveillance.

\subsection{Adaptive Sensing by Reinforcement Learning}
Different from single-pixel camera, which can only adjust the 2D spatial coding patterns, in SCI, our modulation is HD. 
Though in the above subsection, we mentioned that mask optimization may only lead to limited gain, we do have more freedom for adaptive sensing in SCI~\cite{Spinoulas_15_Opt}.
For example, in video SCI, we can adapt the sensing rate by changing the number of patterns during one exposure time. Intuitively, a high-speed scene with complicated objects needs a small Cr ($N_t$) while a low-speed scene with simple objects can have a high Cr.   
A simple look-up table was proposed in~\cite{Yuan13ICIP} to implement this temporal adaptive sensing. 
One near-future research direction is to build a closed-loop adaptive sensing system using SCI and reinforcement learning, which can also integrate with the deep learning-based reconstruction to perform real-time control of SCI.
A similar idea can also be used in spectral SCI, where we can adapt the number of spectral channels depending on the specific task, and other SCI systems. 
Towards this end, a hybrid system of SCI sensing, deep learning reconstruction and reinforcement learning control will be built. This leads to the following future of SCI by collaboration.

\section{Win-Win Future by Collaboration}
In this article, our SCI system only considers the sensing part, \ie, capturing the HD information. 
As mentioned in the beginning of this article, in our daily life, sensing and capturing the data around us is the primary step to perceive the environment, and the main goal is cognition and interaction with the environment. 
Therefore, a joint end-to-end capture, sensing, control, transmission and recognition intelligence system is desired. 
Significant progress has been made in image/video compression, communication, and pattern recognition during the past decade. 
SCI systems, when they are integrated with these systems will lead to a revolution for machine vision.
For instance, every SCI system itself is an encoder and the conventional compression method based on clean images can be changed to coding based on compressed measurements. Meanwhile, the detection and recognition algorithms can be directly performed on the captured compressed measurements~\cite{Lu20SEC}\footnote{Code at: \url{https://www.thecarlab.org/outcomes/software}.} (\eg, an SCI-event camera can be built).   
These revolutionary modules can be re-organized to a new regime of optical compression, coding on measurements, semantic and contextual analysis on measurements, decoding (and refined recognition) on the edge/cloud, and reinforcement learning control.  
This new end-to-end closed-loop integration system resembles evolutionary learning or even further the neural processing and thus the cognition of human beings, which is also the long term research goal of artificial intelligence.

In the future, we envision more and more daily life applications use SCI as the sensing part and integrate with other systems for specific use cases. 

\section*{Acknowledgements}
The authors thank Ziyi Meng for assistance in preparing some figures.

\bibliographystyle{IEEEtran}
\bibliography{sci_spm}

\begin{thebibliography}{100}
\providecommand{\url}[1]{#1}
\csname url@samestyle\endcsname
\providecommand{\newblock}{\relax}
\providecommand{\bibinfo}[2]{#2}
\providecommand{\BIBentrySTDinterwordspacing}{\spaceskip=0pt\relax}
\providecommand{\BIBentryALTinterwordstretchfactor}{4}
\providecommand{\BIBentryALTinterwordspacing}{\spaceskip=\fontdimen2\font plus
\BIBentryALTinterwordstretchfactor\fontdimen3\font minus
  \fontdimen4\font\relax}
\providecommand{\BIBforeignlanguage}[2]{{%
\expandafter\ifx\csname l@#1\endcsname\relax
\typeout{** WARNING: IEEEtran.bst: No hyphenation pattern has been}%
\typeout{** loaded for the language `#1'. Using the pattern for}%
\typeout{** the default language instead.}%
\else
\language=\csname l@#1\endcsname
\fi
#2}}
\providecommand{\BIBdecl}{\relax}
\BIBdecl

\bibitem{Candes06_Robust}
E.~J. {Candes}, J.~{Romberg}, and T.~{Tao}, ``Robust uncertainty principles:
  exact signal reconstruction from highly incomplete frequency information,''
  \emph{IEEE Transactions on Information Theory}, vol.~52, no.~2, pp. 489--509,
  Feb 2006.

\bibitem{Donoho06_CS}
D.~L. {Donoho}, ``Compressed sensing,'' \emph{IEEE Transactions on Information
  Theory}, vol.~52, no.~4, pp. 1289--1306, April 2006.

\bibitem{Gehm07_DDCASSI}
M.~E. Gehm, R.~John, D.~J. Brady, R.~M. Willett, and T.~J. Schulz,
  ``Single-shot compressive spectral imaging with a dual-disperser
  architecture,'' \emph{Opt. Express}, vol.~15, no.~21, pp. 14\,013--14\,027,
  Oct 2007.

\bibitem{Wagadarikar08_CASSI}
A.~Wagadarikar, R.~John, R.~Willett, and D.~Brady, ``Single disperser design
  for coded aperture snapshot spectral imaging,'' \emph{Appl. Opt.}, vol.~47,
  no.~10, pp. B44--B51, Apr 2008.

\bibitem{Wagadarikar09_vCASSI}
A.~A. Wagadarikar, N.~P. Pitsianis, X.~Sun, and D.~J. Brady, ``Video rate
  spectral imaging using a coded aperture snapshot spectral imager,''
  \emph{Opt. Express}, vol.~17, no.~8, pp. 6368--6388, Apr 2009.

\bibitem{Lin14_ACMTG}
X.~Lin, Y.~Liu, J.~Wu, and Q.~Dai, ``Spatial-spectral encoded compressive
  hyperspectral imaging,'' \emph{ACM Trans. Graph.}, vol.~33, no.~6, Nov. 2014.

\bibitem{Yuan15_JSTSP_CASSI_SI}
X.~{Yuan}, T.~{Tsai}, R.~{Zhu}, P.~{Llull}, D.~{Brady}, and L.~{Carin},
  ``Compressive hyperspectral imaging with side information,'' \emph{IEEE
  Journal of Selected Topics in Signal Processing}, vol.~9, no.~6, pp.
  964--976, Sep. 2015.

\bibitem{Hitomi11_ICCV_videoCS}
Y.~{Hitomi}, J.~{Gu}, M.~{Gupta}, T.~{Mitsunaga}, and S.~K. {Nayar}, ``Video
  from a single coded exposure photograph using a learned over-complete
  dictionary,'' in \emph{2011 International Conference on Computer Vision}, Nov
  2011, pp. 287--294.

\bibitem{Reddy11_CVPR_P2C2}
D.~{Reddy}, A.~{Veeraraghavan}, and R.~{Chellappa}, ``P2c2: Programmable pixel
  compressive camera for high speed imaging,'' in \emph{CVPR 2011}, June 2011,
  pp. 329--336.

\bibitem{Llull13_OE_CACTI}
P.~Llull, X.~Liao, X.~Yuan, J.~Yang, D.~Kittle, L.~Carin, G.~Sapiro, and D.~J.
  Brady, ``Coded aperture compressive temporal imaging,'' \emph{Opt. Express},
  vol.~21, no.~9, pp. 10\,526--10\,545, May 2013.

\bibitem{Gao14_Nature_fast}
L.~Gao, J.~Liang, C.~Li, and L.~V. Wang, ``Single-shot compressed ultrafast
  photography at one hundred billion frames per second,'' \emph{Nature}, vol.
  516, no. 7529, pp. 74--77, 2014.

\bibitem{Yuan16BOE}
X.~Yuan and S.~Pang, ``Structured illumination temporal compressive
  microscopy,'' \emph{Biomedical Optics Express}, vol.~7, pp. 746--758, 2016.

\bibitem{brady2015compressive}
D.~J. Brady, A.~Mrozack, K.~MacCabe, and P.~Llull, ``Compressive tomography,''
  \emph{Advances in Optics and Photonics}, vol.~7, no.~4, pp. 756--813, 2015.

\bibitem{Qiao19_OCT}
M.~Qiao, Y.~Sun, X.~Liu, X.~Yuan, and P.~Wilford, ``Snapshot optical coherence
  tomography,'' in \emph{Digital Holography and Three-Dimensional Imaging
  2019}.\hskip 1em plus 0.5em minus 0.4em\relax Optical Society of America,
  2019, p. W4B.3.

\bibitem{Brady09_ComHolo}
D.~J. Brady, K.~Choi, D.~L. Marks, R.~Horisaki, and S.~Lim, ``Compressive
  holography,'' \emph{Opt. Express}, vol.~17, no.~15, pp. 13\,040--13\,049, Jul
  2009.

\bibitem{Brady18_CPRL_omHolo}
W.~Zhang, L.~Cao, D.~J. Brady, H.~Zhang, J.~Cang, H.~Zhang, and G.~Jin,
  ``Twin-image-free holography: A compressive sensing approach,'' \emph{Phys.
  Rev. Lett.}, vol. 121, p. 093902, Aug 2018.

\bibitem{Wang17_OE_HoloVideo}
Z.~Wang, L.~Spinoulas, K.~He, L.~Tian, O.~Cossairt, A.~K. Katsaggelos, and
  H.~Chen, ``Compressive holographic video,'' \emph{Opt. Express}, vol.~25,
  no.~1, pp. 250--262, Jan 2017.

\bibitem{He20_Lightfield_OE}
K.~He, X.~Wang, Z.~W. Wang, H.~Yi, N.~F. Scherer, A.~K. Katsaggelos, and
  O.~Cossairt, ``Snapshot multifocal light field microscopy,'' \emph{Opt.
  Express}, vol.~28, no.~8, pp. 12\,108--12\,120, Apr 2020.

\bibitem{Tsai15OE}
T.-H. Tsai, X.~Yuan, and D.~J. Brady, ``Spatial light modulator based color
  polarization imaging,'' \emph{Optics Express}, vol.~23, no.~9, pp.
  11\,912--11\,926, May 2015.

\bibitem{Nayar_2000_CVPR_HDR}
S.~K. {Nayar} and T.~{Mitsunaga}, ``High dynamic range imaging: spatially
  varying pixel exposures,'' in \emph{Proceedings IEEE Conference on Computer
  Vision and Pattern Recognition. CVPR 2000}, vol.~1, no.~1, June 2000, pp.
  472--479.

\bibitem{Nayar_03_ICCV_HDRSLM}
S.~K. {Nayar} and V.~{Branzoi}, ``Adaptive dynamic range imaging: optical
  control of pixel exposures over space and time,'' in \emph{Proceedings Ninth
  IEEE International Conference on Computer Vision}, vol.~2, Oct 2003, pp.
  1168--1175.

\bibitem{Yuan_16_OE}
X.~Yuan, ``Compressive dynamic range imaging via {Bayesian} shrinkage
  dictionary learning,'' \emph{Optical Engineering}, vol.~55, no.~12, p.
  123110, 2016.

\bibitem{Llull15Optica}
P.~Llull, X.~Yuan, L.~Carin, and D.~Brady, ``Image translation for single-shot
  focal tomography,'' \emph{Optica}, vol.~2, no.~9, pp. 822--825, 2015.

\bibitem{Tsai15OL}
T.-H. Tsai, P.~Llull, X.~Yuan, D.~J. Brady, and L.~Carin, ``Spectral-temporal
  compressive imaging,'' \emph{Optics Letters}, vol.~40, no.~17, pp.
  4054--4057, Sep 2015.

\bibitem{Tsai13_AO_SpecPol}
T.-H. Tsai and D.~J. Brady, ``Coded aperture snapshot spectral polarization
  imaging,'' \emph{Appl. Opt.}, vol.~52, no.~10, pp. 2153--2161, Apr 2013.

\bibitem{Qiao2020_CACTI}
M.~Qiao, X.~Liu, and X.~Yuan, ``Snapshot spatial--temporal compressive
  imaging,'' \emph{Opt. Lett.}, vol.~45, no.~7, pp. 1659--1662, Apr 2020.

\bibitem{Sun16OE}
Y.~Sun, X.~Yuan, and S.~Pang, ``High-speed compressive range imaging based on
  active illumination,'' \emph{Optics Express}, vol.~24, no.~20, pp.
  22\,836--22\,846, Oct 2016.

\bibitem{Sun17OE}
------, ``Compressive high-speed stereo imaging,'' \emph{Opt Express}, vol.~25,
  no.~15, pp. 18\,182--18\,190, 2017.

\bibitem{Yuan14CVPR}
X.~Yuan, P.~Llull, X.~Liao, J.~Yang, D.~J. Brady, G.~Sapiro, and L.~Carin,
  ``Low-cost compressive sensing for color video and depth,'' in \emph{IEEE
  Conference on Computer Vision and Pattern Recognition (CVPR)}, 2014, pp.
  3318--3325.

\bibitem{Bioucas-Dias2007TwIST}
J.~Bioucas-Dias and M.~Figueiredo, ``A new {TwIST}: Two-step iterative
  shrinkage/thresholding algorithms for image restoration,'' \emph{IEEE
  Transactions on Image Processing}, vol.~16, no.~12, pp. 2992--3004, December
  2007.

\bibitem{Yuan16ICIP_GAP}
X.~Yuan, ``Generalized alternating projection based total variation
  minimization for compressive sensing,'' in \emph{2016 IEEE International
  Conference on Image Processing (ICIP)}, Sept 2016, pp. 2539--2543.

\bibitem{Yang14GMM}
J.~Yang, X.~Yuan, X.~Liao, P.~Llull, G.~Sapiro, D.~J. Brady, and L.~Carin,
  ``Video compressive sensing using {G}aussian mixture models,'' \emph{IEEE
  Transaction on Image Processing}, vol.~23, no.~11, pp. 4863--4878, November
  2014.

\bibitem{Yang14GMMonline}
J.~Yang, X.~Liao, X.~Yuan, P.~Llull, D.~J. Brady, G.~Sapiro, and L.~Carin,
  ``Compressive sensing by learning a {G}aussian mixture model from
  measurements,'' \emph{IEEE Transaction on Image Processing}, vol.~24, no.~1,
  pp. 106--119, January 2015.

\bibitem{Rajwade13}
A.~Rajwade, D.~Kittle, T.-H. Tsai, and L.~Carin, ``Coded hyperspectral imaging
  and blind compressive sensing,'' \emph{SIAM Journal on Imaging Sciences},
  vol.~6, no.~2, pp. 782--812, 2013.

\bibitem{Yuan16AO}
X.~Yuan, X.~Liao, P.~Llull, D.~Brady, and L.~Carin, ``Efficient patch-based
  approach for compressive depth imaging,'' \emph{Applied Optics}, vol.~55,
  no.~27, pp. 7556--7564, Sep 2016.

\bibitem{Yuan2020_CVPR_PnP}
X.~{Yuan}, Y.~{Liu}, J.~{Suo}, and Q.~{Dai}, ``Plug-and-play algorithms for
  large-scale snapshot compressive imaging,'' in \emph{IEEE/CVF Conference on
  Computer Vision and Pattern Recognition (CVPR)}, June 2020.

\bibitem{Qiao2020_APLP}
M.~Qiao, Z.~Meng, J.~Ma, and X.~Yuan, ``Deep learning for video compressive
  sensing,'' \emph{APL Photonics}, vol.~5, no.~3, p. 030801, 2020.

\bibitem{Mait18_AOP_CI}
J.~N. Mait, G.~W. Euliss, and R.~A. Athale, ``Computational imaging,''
  \emph{Adv. Opt. Photon.}, vol.~10, no.~2, pp. 409--483, Jun 2018.

\bibitem{Altmann18Science}
Y.~Altmann, S.~McLaughlin, M.~J. Padgett, V.~K. Goyal, A.~O. Hero, and
  D.~Faccio, ``Quantum-inspired computational imaging,'' \emph{Science}, vol.
  361, no. 6403, 2018.

\bibitem{Duarte08SPM}
M.~F. Duarte, M.~A. Davenport, D.~Takhar, J.~N. Laska, T.~Sun, K.~F. Kelly, and
  R.~G. Baraniuk, ``Single-pixel imaging via compressive sampling,'' \emph{IEEE
  Signal Processing Magazine}, vol.~25, no.~2, pp. 83--91, 2008.

\bibitem{lustig2007sparse}
M.~Lustig, D.~Donoho, and J.~M. Pauly, ``Sparse mri: The application of
  compressed sensing for rapid mr imaging,'' \emph{Magnetic Resonance in
  Medicine: An Official Journal of the International Society for Magnetic
  Resonance in Medicine}, vol.~58, no.~6, pp. 1182--1195, 2007.

\bibitem{Jalali19TIT_SCI}
S.~{Jalali} and X.~{Yuan}, ``Snapshot compressed sensing: Performance bounds
  and algorithms,'' \emph{IEEE Transactions on Information Theory}, vol.~65,
  no.~12, pp. 8005--8024, Dec 2019.

\bibitem{Iliadis18DSPvideoCS}
M.~Iliadis, L.~Spinoulas, and A.~K. Katsaggelos, ``Deep fully-connected
  networks for video compressive sensing,'' \emph{Digital Signal Processing},
  vol.~72, pp. 9--18, 2018.

\bibitem{Lucas18SPM}
A.~{Lucas}, M.~{Iliadis}, R.~{Molina}, and A.~K. {Katsaggelos}, ``Using deep
  neural networks for inverse problems in imaging: Beyond analytical methods,''
  \emph{IEEE Signal Processing Magazine}, vol.~35, no.~1, pp. 20--36, Jan 2018.

\bibitem{Ma19ICCV}
J.~Ma, X.~Liu, Z.~Shou, and X.~Yuan, ``Deep tensor admm-net for snapshot
  compressive imaging,'' in \emph{IEEE/CVF Conference on Computer Vision
  (ICCV)}, 2019.

\bibitem{Miao19ICCV}
X.~Miao, X.~Yuan, Y.~Pu, and V.~Athitsos, ``$\lambda$-net: Reconstruct
  hyperspectral images from a snapshot measurement,'' in \emph{IEEE/CVF
  Conference on Computer Vision (ICCV)}, 2019.

\bibitem{Iliadis_DSP2020}
M.~Iliadis, L.~Spinoulas, and A.~K. Katsaggelos, ``Deepbinarymask: Learning a
  binary mask for video compressive sensing,'' \emph{Digital Signal
  Processing}, vol.~96, p. 102591, 2020.

\bibitem{Meng2020_OL_SHEM}
Z.~Meng, M.~Qiao, J.~Ma, Z.~Yu, K.~Xu, and X.~Yuan, ``Snapshot multispectral
  endomicroscopy,'' \emph{Opt. Lett.}, vol.~45, no.~14, pp. 3897--3900, Jul
  2020.

\bibitem{Meng20ECCV_TSAnet}
Z.~Meng, J.~Ma, and X.~Yuan, ``End-to-end low cost compressive spectral imaging
  with spatial-spectral self-attention,'' in \emph{European Conference on
  Computer Vision (ECCV)}, August 2020.

\bibitem{Cheng20ECCV_Birnat}
Z.~Cheng, R.~Lu, Z.~Wang, H.~Zhang, B.~Chen, Z.~Meng, and X.~Yuan, ``{BIRNAT}:
  Bidirectional recurrent neural networks with adversarial training for video
  snapshot compressive imaging,'' in \emph{European Conference on Computer
  Vision (ECCV)}, August 2020.

\bibitem{Cao2011_Cao_Prism}
X.~{Cao}, H.~{Du}, X.~{Tong}, Q.~{Dai}, and S.~{Lin}, ``A prism-mask system for
  multispectral video acquisition,'' \emph{IEEE Transactions on Pattern
  Analysis and Machine Intelligence}, vol.~33, no.~12, pp. 2423--2435, Dec
  2011.

\bibitem{Liu_2016_GISR}
Z.~Liu, S.~Tan, J.~Wu, E.~Li, X.~Shen, and S.~Han, ``Spectral camera based on
  ghost imaging via sparsity constraints,'' \emph{Sci Rep.}, vol. 2016, no.~6,
  p. 25718, 2016.

\bibitem{Koller15_videoCS}
R.~Koller, L.~Schmid, N.~Matsuda, T.~Niederberger, L.~Spinoulas, O.~Cossairt,
  G.~Schuster, and A.~K. Katsaggelos, ``High spatio-temporal resolution video
  with compressed sensing,'' \emph{Opt. Express}, vol.~23, no.~12, pp.
  15\,992--16\,007, Jun 2015.

\bibitem{Llull14_COSI_focalstack}
P.~Llull, X.~Yuan, X.~Liao, J.~Yang, L.~Carin, G.~Sapiro, and D.~J. Brady,
  ``Compressive extended depth of field using image space coding,'' in
  \emph{Classical Optics 2014}.\hskip 1em plus 0.5em minus 0.4em\relax Optical
  Society of America, 2014, p. CM2D.3.

\bibitem{duarte2019computed}
J.~Duarte, R.~Cassin, J.~Huijts, B.~Iwan, F.~Fortuna, L.~Delbecq, H.~Chapman,
  M.~Fajardo, M.~Kovacev, W.~Boutu \emph{et~al.}, ``Computed stereo lensless
  x-ray imaging,'' \emph{Nature Photonics}, vol.~13, no.~7, pp. 449--453, 2019.

\bibitem{Antipa18_Optica_3D}
N.~Antipa, G.~Kuo, R.~Heckel, B.~Mildenhall, E.~Bostan, R.~Ng, and L.~Waller,
  ``Diffusercam: lensless single-exposure 3d imaging,'' \emph{Optica}, vol.~5,
  no.~1, pp. 1--9, Jan 2018.

\bibitem{Qiao_CSOCT_arxiv2020}
M.~Qiao, Y.~Sun, J.~Ma, Z.~Meng, X.~Liu, and X.~Yuan, ``Snapshot
  interferometric 3d imaging by compressive sensing and deep learning,''
  \emph{arXiv: 2004.02633}, April 2020.

\bibitem{Brady13_xray}
D.~J. Brady, D.~L. Marks, K.~P. MacCabe, and J.~A. O'Sullivan, ``Coded
  apertures for x-ray scatter imaging,'' \emph{Appl. Opt.}, vol.~52, no.~32,
  pp. 7745--7754, Nov 2013.

\bibitem{Greenberg:13_xray}
J.~A. Greenberg, K.~Krishnamurthy, and D.~Brady, ``Snapshot molecular imaging
  using coded energy-sensitive detection,'' \emph{Opt. Express}, vol.~21,
  no.~21, pp. 25\,480--25\,491, Oct 2013.

\bibitem{Hassan16_xray}
M.~Hassan, J.~A. Greenberg, I.~Odinaka, and D.~J. Brady, ``Snapshot fan beam
  coded aperture coherent scatter tomography,'' \emph{Opt. Express}, vol.~24,
  no.~16, pp. 18\,277--18\,289, Aug 2016.

\bibitem{MacCabe13_xray}
K.~P. MacCabe, A.~D. Holmgren, M.~P. Tornai, and D.~J. Brady, ``Snapshot 2d
  tomography via coded aperture x-ray scatter imaging,'' \emph{Appl. Opt.},
  vol.~52, no.~19, pp. 4582--4589, Jul 2013.

\bibitem{Zhu:18_xray}
Z.~Zhu, R.~A. Ellis, and S.~Pang, ``Coded cone-beam x-ray diffraction
  tomography with a low-brilliance tabletop source,'' \emph{Optica}, vol.~5,
  no.~6, pp. 733--738, Jun 2018.

\bibitem{Babacan12_lightfield}
S.~D. {Babacan}, R.~{Ansorge}, M.~{Luessi}, P.~R. {Mataran}, R.~{Molina}, and
  A.~K. {Katsaggelos}, ``Compressive light field sensing,'' \emph{IEEE
  Transactions on Image Processing}, vol.~21, no.~12, pp. 4746--4757, 2012.

\bibitem{Ruiz13_lightfield}
P.~{Ruiz}, J.~{Mateos}, M.~C. {Cárdenas}, S.~{Nakajima}, R.~{Molina}, and
  A.~K. {Katsaggelos}, ``Light field acquisition from blurred observations
  using a programmable coded aperture camera,'' in \emph{21st European Signal
  Processing Conference (EUSIPCO 2013)}, 2013, pp. 1--5.

\bibitem{Portz13_ICCP_hdrt}
T.~{Portz}, L.~{Zhang}, and H.~{Jiang}, ``Random coded sampling for high-speed
  hdr video,'' in \emph{IEEE International Conference on Computational
  Photography (ICCP)}, April 2013, pp. 1--8.

\bibitem{Ma2021_LeSTI_OE}
X.~Ma, X.~Yuan, C.~Fu, and G.~R. Arce, ``Led-based compressive spectral
  temporal imaging system,'' \emph{Optics Express}, 2021.

\bibitem{Pang14_xray}
S.~Pang, M.~Hassan, J.~Greenberg, A.~Holmgren, K.~Krishnamurthy, and D.~Brady,
  ``Complementary coded apertures for 4-dimensional x-ray coherent scatter
  imaging,'' \emph{Opt. Express}, vol.~22, no.~19, pp. 22\,925--22\,936, Sep
  2014.

\bibitem{Gopalsami12_Hadamard}
N.~Gopalsami, S.~Liao, T.~Elmer, E.~Koehl, A.~Heifetz, A.~Raptis, L.~Spinoulas,
  and A.~Katsaggelos, ``Passive millimeter-wave imaging with compressive
  sensing,'' vol.~51, no.~9, 2012.

\bibitem{Spinoulas:12_hadamard}
L.~Spinoulas, J.~Qi, A.~K. Katsaggelos, T.~W. Elmer, N.~Gopalsami, and A.~C.
  Raptis, ``Optimized compressive sampling for passive millimeter-wave
  imaging,'' \emph{Appl. Opt.}, vol.~51, no.~26, pp. 6335--6342, Sep 2012.

\bibitem{Yuan2020TMM}
X.~{Yuan} and R.~{Haimi-Cohen}, ``Image compression based on compressive
  sensing: End-to-end comparison with jpeg,'' \emph{IEEE Transactions on
  Multimedia}, vol.~22, no.~11, pp. 2889--2904, 2020.

\bibitem{Tsiligianni14_TIT}
E.~V. {Tsiligianni}, L.~P. {Kondi}, and A.~K. {Katsaggelos}, ``Construction of
  incoherent unit norm tight frames with application to compressed sensing,''
  \emph{IEEE Transactions on Information Theory}, vol.~60, no.~4, pp.
  2319--2330, 2014.

\bibitem{Liu19_PAMI_DeSCI}
Y.~{Liu}, X.~{Yuan}, J.~{Suo}, D.~J. {Brady}, and Q.~{Dai}, ``Rank minimization
  for snapshot compressive imaging,'' \emph{IEEE Transactions on Pattern
  Analysis and Machine Intelligence}, vol.~41, no.~12, pp. 2990--3006, Dec
  2019.

\bibitem{jalali2016compression}
S.~Jalali and A.~Maleki, ``From compression to compressed sensing,''
  \emph{Applied and Computational Harmonic Analysis}, vol.~40, no.~2, pp.
  352--385, 2016.

\bibitem{Wang15_SIAM}
L.~Wang, J.~Huang, X.~Yuan, K.~Krishnamurthy, J.~Greenberg, V.~Cevher, M.~R.~D.
  Rodrigues, D.~Brady, R.~Calderbank, and L.~Carin, ``Signal recovery and
  system calibration from multiple compressive poisson measurements,''
  \emph{SIAM Journal on Imaging Sciences}, vol.~8, no.~3, pp. 1923--1954, 2015.

\bibitem{Yuan17AO}
X.~Yuan, Y.~Sun, and S.~Pang, ``Compressive video sensing with side
  information,'' \emph{Appl. Opt.}, vol.~56, no.~10, pp. 2697--2704, 2017.

\bibitem{Figueiredo07GPSR}
M.~A.~T. Figueiredo, R.~D. Nowak, and S.~J. Wright, ``Gradient projection for
  sparse reconstruction: Application to compressed sensing and other inverse
  problems,'' \emph{IEEE Journal of Selected Topics in Signal Processing},
  vol.~1, no.~4, pp. 586--597, Dec 2007.

\bibitem{Wang17_PAMI_GSC}
L.~{Wang}, Z.~{Xiong}, G.~{Shi}, F.~{Wu}, and W.~{Zeng}, ``Adaptive nonlocal
  sparse representation for dual-camera compressive hyperspectral imaging,''
  \emph{IEEE Transactions on Pattern Analysis and Machine Intelligence},
  vol.~39, no.~10, pp. 2104--2111, Oct 2017.

\bibitem{Xiong_17_CVPRW_HSCNN}
Z.~{Xiong}, Z.~{Shi}, H.~{Li}, L.~{Wang}, D.~{Liu}, and F.~{Wu}, ``Hscnn:
  Cnn-based hyperspectral image recovery from spectrally undersampled
  projections,'' in \emph{2017 IEEE International Conference on Computer Vision
  Workshops (ICCVW)}, Oct 2017, pp. 518--525.

\bibitem{Wang19_CVPR_HSSP}
L.~{Wang}, C.~{Sun}, Y.~{Fu}, M.~H. {Kim}, and H.~{Huang}, ``Hyperspectral
  image reconstruction using a deep spatial-spectral prior,'' in \emph{2019
  IEEE/CVF Conference on Computer Vision and Pattern Recognition (CVPR)}, June
  2019, pp. 8024--8033.

\bibitem{Yang20_TIP_SeSCI}
P.~{Yang}, L.~{Kong}, X.~{Liu}, X.~{Yuan}, and G.~{Chen}, ``Shearlet enhanced
  snapshot compressive imaging,'' \emph{IEEE Transactions on Image Processing},
  vol.~29, pp. 6466--6481, 2020.

\bibitem{Yuan_TVSCI_arxiv2020}
X.~Yuan, ``Various total variation for snapshot video compressive imaging,''
  \emph{arXiv: 2005.08028}, May 2020.

\bibitem{Boyd11ADMM}
S.~Boyd, N.~Parikh, E.~Chu, B.~Peleato, and J.~Eckstein, ``Distributed
  optimization and statistical learning via the alternating direction method of
  multipliers,'' \emph{Foundations and Trends in Machine Learning}, vol.~3,
  no.~1, pp. 1--122, January 2011.

\bibitem{Yuan16SJ}
X.~Yuan, H.~Jiang, G.~Huang, and P.~Wilford, ``{SLOPE}: Shrinkage of local
  overlapping patches estimator for lensless compressive imaging,'' \emph{IEEE
  Sensors Journal}, vol.~16, no.~22, pp. 8091--8102, November 2016.

\bibitem{Mertzler14Denoising}
C.~A. Metzler, A.~Maleki, and R.~G. Baraniuk, ``From denoising to compressed
  sensing,'' \emph{IEEE Transactions on Information Theory}, vol.~62, no.~9,
  pp. 5117--5144, 2016.

\bibitem{Chan2017PlugandPlayAF}
S.~H. Chan, X.~Wang, and O.~A. Elgendy, ``Plug-and-play {ADMM} for image
  restoration: Fixed-point convergence and applications,'' \emph{IEEE
  Transactions on Computational Imaging}, vol.~3, pp. 84--98, 2017.

\bibitem{Liao14GAP}
X.~Liao, H.~Li, and L.~Carin, ``Generalized alternating projection for
  weighted-$\ell_{2,1}$ minimization with applications to model-based
  compressive sensing,'' \emph{SIAM Journal on Imaging Sciences}, vol.~7,
  no.~2, pp. 797--823, 2014.

\bibitem{Gu14CVPR}
S.~Gu, L.~Zhang, W.~Zuo, and X.~Feng, ``Weighted nuclear norm minimization with
  application to image denoising,'' in \emph{IEEE Conference on Computer Vision
  and Pattern Recognition (CVPR)}, 2014, pp. 2862--2869.

\bibitem{Renna16_GMM}
F.~{Renna}, L.~{Wang}, X.~{Yuan}, J.~{Yang}, G.~{Reeves}, R.~{Calderbank},
  L.~{Carin}, and M.~R.~D. {Rodrigues}, ``Classification and reconstruction of
  high-dimensional signals from low-dimensional features in the presence of
  side information,'' \emph{IEEE Transactions on Information Theory}, vol.~62,
  no.~11, pp. 6459--6492, 2016.

\bibitem{Aharon06TSP}
M.~Aharon, M.~Elad, and A.~Bruckstein, ``{K-SVD}: An algorithm for designing
  overcomplete dictionaries for sparse representation,'' \emph{IEEE
  Transactions on Signal Processing}, vol.~54, no.~11, pp. 4311--4322, 2006.

\bibitem{Chen10_TSP_GMM}
M.~{Chen}, J.~{Silva}, J.~{Paisley}, C.~{Wang}, D.~{Dunson}, and L.~{Carin},
  ``Compressive sensing on manifolds using a nonparametric mixture of factor
  analyzers: Algorithm and performance bounds,'' \emph{IEEE Transactions on
  Signal Processing}, vol.~58, no.~12, pp. 6140--6155, 2010.

\bibitem{Mairal_ICCV09_NLSC}
J.~{Mairal}, F.~{Bach}, J.~{Ponce}, G.~{Sapiro}, and A.~{Zisserman},
  ``Non-local sparse models for image restoration,'' in \emph{2009 IEEE 12th
  International Conference on Computer Vision}, Sep. 2009, pp. 2272--2279.

\bibitem{Zha2020_TIP_BenchSC}
Z.~{Zha}, X.~{Yuan}, B.~{Wen}, J.~{Zhou}, J.~{Zhang}, and C.~{Zhu}, ``A
  benchmark for sparse coding: When group sparsity meets rank minimization,''
  \emph{IEEE Transactions on Image Processing}, vol.~29, pp. 5094--5109, 2020.

\bibitem{Barbastathis19DL}
G.~Barbastathis, A.~Ozcan, and G.~Situ, ``On the use of deep learning for
  computational imaging,'' \emph{Optica}, vol.~6, no.~8, pp. 921--943, Aug
  2019.

\bibitem{Unet_RFB15a}
O.~Ronneberger, P.~Fischer, and T.~Brox, ``U-net: Convolutional networks for
  biomedical image segmentation,'' in \emph{Medical Image Computing and
  Computer-Assisted Intervention (MICCAI)}, ser. LNCS, vol. 9351.\hskip 1em
  plus 0.5em minus 0.4em\relax Springer, 2015, pp. 234--241.

\bibitem{Goodfellow14_GAN}
I.~Goodfellow, J.~Pouget-Abadie, M.~Mirza, B.~Xu, D.~Warde-Farley, S.~Ozair,
  A.~Courville, and Y.~Bengio, ``Generative adversarial nets,'' in
  \emph{Advances in Neural Information Processing Systems 27}, 2014, pp.
  2672--2680.

\bibitem{Vaswani17_Attention}
A.~Vaswani, N.~Shazeer, N.~Parmar, J.~Uszkoreit, L.~Jones, A.~N. Gomez, L.~u.
  Kaiser, and I.~Polosukhin, ``Attention is all you need,'' in \emph{Advances
  in Neural Information Processing Systems 30}, 2017, pp. 5998--6008.

\bibitem{Johnson2016Perceptual}
J.~Johnson, A.~Alahi, and L.~Fei-Fei, ``Perceptual losses for real-time style
  transfer and super-resolution,'' in \emph{European Conference on Computer
  Vision}, 2016.

\bibitem{hershey2014deep_unfold}
J.~R. Hershey, J.~L. Roux, and F.~Weninger, ``Deep unfolding: Model-based
  inspiration of novel deep architectures,'' \emph{arXiv preprint
  arXiv:1409.2574}, 2014.

\bibitem{Yang_NIP16_ADMM-net}
Y.~Yang, J.~Sun, H.~Li, and Z.~Xu, ``Deep admm-net for compressive sensing
  mri,'' in \emph{Advances in Neural Information Processing Systems 29}, 2016,
  pp. 10--18.

\bibitem{Meng_GAPnet_arxiv2020}
Z.~Meng, S.~Jalali, and X.~Yuan, ``Gap-net for snapshot compressive imaging,''
  \emph{arXiv: 2012.08364}, December 2020.

\bibitem{Huang2021_CVPR_GSMSCI}
T.~{Huang}, W.~Dong, X.~{Yuan}, J.~Wu, , and G.~{Shi}, ``Deep gaussian scale
  mixture prior for spectral compressive imaging,'' in \emph{IEEE/CVF
  Conference on Computer Vision and Pattern Recognition (CVPR)}, June 2021.

\bibitem{Zhang18TIP_FFDNet}
K.~Zhang, W.~Zuo, and L.~Zhang, ``{FFDNet}: Toward a fast and flexible solution
  for {CNN}-based image denoising,'' \emph{IEEE Trans. Image Processing},
  vol.~27, no.~9, pp. 4608--4622, 2018.

\bibitem{Ryu2019PlugandPlayMP}
E.~K. Ryu, J.~Liu, S.~Wang, X.~Chen, Z.~Wang, and W.~Yin, ``Plug-and-play
  methods provably converge with properly trained denoisers,'' in
  \emph{International Conference on Machine Learning (ICML)}, 2019.

\bibitem{PnP_SCI_arxiv2021}
X.~Yuan, J.~S. Yang~Liu, F.~Durand, and Q.~Dai, ``Plug-and-play algorithms for
  video snapshot compressive imaging,'' \emph{arXiv: 2101.04822}, Jan 2021.

\bibitem{Zheng20_PRJ_PnP-CASSI}
S.~Zheng, Y.~Liu, Z.~Meng, M.~Qiao, Z.~Tong, X.~Yang, S.~Han, and X.~Yuan,
  ``Deep plug-and-play priors for spectral snapshot compressive imaging,''
  \emph{Photon. Res.}, vol.~9, no.~2, pp. B18--B29, Feb 2021.

\bibitem{Wang2021_CVPR_MetaSCI}
Z.~{Wang}, H.~{Zhang}, Z.~{Cheng}, B.~Chen, and X.~{Yuan}, ``Metasci: Scalable
  and adaptive reconstruction for video compressive sensing,'' in
  \emph{IEEE/CVF Conference on Computer Vision and Pattern Recognition (CVPR)},
  June 2021.

\bibitem{Cheng2021_CVPR_ReverSCI}
Z.~{Cheng}, B.~Chen, G.~{Liu}, H.~Zhang, R.~{Lu}, Z.~Wang, and X.~{Yuan},
  ``Memory-efficient network for large-scale video compressive sensing,'' in
  \emph{IEEE/CVF Conference on Computer Vision and Pattern Recognition (CVPR)},
  June 2021.

\bibitem{Lempitsky2018CVPR_DIP}
V.~{Lempitsky}, A.~{Vedaldi}, and D.~{Ulyanov}, ``Deep image prior,'' in
  \emph{2018 IEEE/CVF Conference on Computer Vision and Pattern Recognition},
  2018, pp. 9446--9454.

\bibitem{Qiao2021_MicroCACTI}
M.~Qiao, X.~Liu, and X.~Yuan, ``Snapshot temporal compressive microscopy using
  an iterative algorithm with untrained neural networks,'' \emph{Opt. Lett.},
  2021.

\bibitem{Wang04imagequality}
Z.~Wang, A.~C. Bovik, H.~R. Sheikh, E.~P. Simoncelli \emph{et~al.}, ``Image
  quality assessment: From error visibility to structural similarity,''
  \emph{IEEE Transactions on Image Processing}, vol.~13, no.~4, pp. 600--612,
  2004.

\bibitem{choi2017high}
I.~Choi, D.~S. Jeon, G.~Nam, D.~Gutierrez, and M.~H. Kim, ``High-quality
  hyperspectral reconstruction using a spectral prior,'' vol.~36, no.~6.\hskip
  1em plus 0.5em minus 0.4em\relax ACM, 2017, p. 218.

\bibitem{Liu_2019PIEEE_Edge}
S.~{Liu}, L.~{Liu}, J.~{Tang}, B.~{Yu}, Y.~{Wang}, and W.~{Shi}, ``Edge
  computing for autonomous driving: Opportunities and challenges,''
  \emph{Proceedings of the IEEE}, vol. 107, no.~8, pp. 1697--1716, 2019.

\bibitem{Cui2014_LSA_meta}
T.~J. Cui, M.~Q. Qi, X.~Wan, J.~Zhou, and Q.~Cheng, ``Coding metamaterials,
  digital metamaterials and programmable metamaterials,'' \emph{Light Sci
  Appl}, vol.~3, pp. 1--9, 2014.

\bibitem{Spinoulas_15_Opt}
L.~{Spinoulas}, O.~{Cossairt}, and A.~K. {Katsaggelos}, ``Sampling optimization
  for on-chip compressive video,'' in \emph{2015 IEEE International Conference
  on Image Processing (ICIP)}, 2015, pp. 3329--3333.

\bibitem{Yuan13ICIP}
X.~Yuan, J.~Yang, P.~Llull, X.~Liao, G.~Sapiro, D.~J. Brady, and L.~Carin,
  ``Adaptive temporal compressive sensing for video,'' in \emph{2013 IEEE
  International Conference on Image Processing (ICIP)}, Sept 2013, pp. 14--18.

\bibitem{Lu20SEC}
S.~{Lu}, X.~{Yuan}, and W.~{Shi}, ``Edge compression: An integrated framework
  for compressive imaging processing on cavs,'' in \emph{2020 IEEE/ACM
  Symposium on Edge Computing (SEC)}, 2020, pp. 125--138.

\end{thebibliography}

%
%
%
%
%
%

\end{document}